\documentclass{article}

\usepackage{arxiv}

\usepackage[utf8]{inputenc} 
\usepackage[T1]{fontenc}    
\usepackage{hyperref}       
\usepackage{url}            
\usepackage{booktabs}       
\usepackage{amsfonts}       
\usepackage{amsmath}
\usepackage{amssymb}
\usepackage{bbold}          
\usepackage{nicefrac}       
\usepackage{microtype}      
\usepackage{graphicx}
\usepackage{natbib}
\usepackage{doi}
\usepackage[usenames,dvipsnames]{color}
\usepackage{xcolor}

\usepackage{listings}

\colorlet{lightorange}{orange!05}
\title{Conditional bootstrap for non-linear mixed effects models}

\date{May 4th, 2026}	

\author{ 
    \href{https://orcid.org/0000-0003-2110-6392}{\includegraphics[scale=0.06]{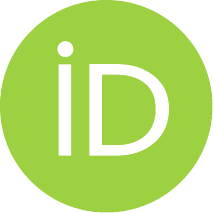}\hspace{1mm}Sofia Kaisaridi} \\
Université Paris Cité \& \\
Université Sorbonne Paris Nord\\ 
Inserm, IAME, F-75018 Paris, France\\
	\And
    \href{https://orcid.org/0000-0002-5709-4322}{\includegraphics[scale=0.06]{orcid.pdf}\hspace{1mm} Moreno Ursino} \\
     Inserm, Inria, Université Paris Cité \\
     UMRS 1346 \\
    F-75015 Paris \\
    France \\
    \And
   \href{https://orcid.org/0000-0002-9150-9886}{\includegraphics[scale=0.06]{orcid.pdf}\hspace{1mm}Emmanuelle Comets}\thanks{Corresponding author} \\
Université Paris Cité \& \\
Université Sorbonne Paris Nord\\ 
Inserm, IAME, F-75018 Paris, France\\
Univ Rennes, Inserm, EHESP \\
Irset UMR\_S 1085 \\
F-35000 Rennes, France \\
	\texttt{emmanuelle.comets@inserm.fr} \\
}

\renewcommand{\headeright}{Conditional bootstrap for NLMEM}
\renewcommand{\undertitle}{{\sf arXiv} preprint}
\renewcommand{\shorttitle}{{\sf arXiv} preprint}

\hypersetup{
pdftitle={conditionalBootstrap},
pdfauthor={Sofia Kaisaridi, Moreno Ursino, Emmanuelle Comets},
pdfkeywords={Non-linear mixed effect models, SAEM algorithm, saemix, R package, parameter estimation, uncertainty, bootstrap, conditional distribution},
}

\begin{document}
\maketitle

\paragraph{Corresponding author:} Emmanuelle Comets \\
\begin{tabular}{p{3.2cm} l}
&INSERM - UMR1137 \\
&UFR de Médecine Site Bichat \\
&16 rue Henri Huchard \\
& 75018 Paris, France \\
&Tel: (33) 6 25 82 49 50\\
&\texttt{emmanuelle.comets@inserm.fr} \\
\end{tabular}

\newpage
\begin{abstract} 
\textit{\bf Background and Objective:} Uncertainty in non-linear mixed effect models is often assessed using the Fisher information matrix to derive the standard errors of estimation. The bootstrap is an alternative to the asymptotic method, with different approaches to handle the different levels of individual and population variabilities. \\
\textit{\bf Methods:} We propose here a non-parametric bootstrap, cNP, resampling from the conditional distribution of the individual parameters, preserving the structure of the original data. cNP was implemented in the saemix package for R along with the case, parametric (Par), and non-parametric (NP) residual bootstraps. Coverage rates were compared in a simulation study using sigmoid Emax models, with rich, sparse and unbalanced designs, and 3 levels of residual variability. \\ 
\textit{\bf Results:} The asymptotic method tended to produce lower than theoretical coverages for the variance terms. Bootstraps provided more adequate coverage, but none of the approaches maintained coverage when the residual error increased. Overall, the new cNP and the Case provided better coverage than the classical NP. \\
\textit{\bf Conclusion:} The new conditional non-parametric bootstrap could be used when it is important to preserve the structure of the original dataset, such as the number of observations or the repartition of covariates as it does not require stratification.
\end{abstract}

\keywords{Non-linear mixed effect models \and SAEM algorithm \and Bootstrap \and Conditional distribution \and Uncertainty \and saemix \and Unbalanced design \and longitudinal data}

\newpage
\section{Introduction} \label{sec:1.Intro}

\hskip 18pt In pharmacological and medical studies multiple observations are generally collected from each individual followed during the trial, for instance drug concentrations in pharmacokinetic (PK) studies, or the evolution of a biomarker linked to disease progression or treatment effect. Hierarchical mixed effects models are used to simultaneously describe the evolution of the underlying physiological and pathological processes and handle different levels of variability, both within observations from the same individual (residual variability) and between individuals (between-subject variability). Drug development makes frequent use of non-linear mixed effects models (NLMEM) to analyse pharmacokinetic and pharmacodynamic data~(\cite{Pillai05}). The results from these models are then used to help design the next stages of clinical decisions and may even influence go/no-go decisions~(\cite{Lalonde07}). Informed decisions need to take into account the probability of success, and therefore incorporate uncertainty at different levels, including the uncertainty associated to parameter estimation. In NLMEM, the standard approach to compute estimation errors is an asymptotic approximation based on the Fisher Information Matrix, the inverse of which has been shown to be a lower bound on the standard errors of estimation (SE) of the model parameters~(\cite{Davidian95}). The validity of this approximation requires that both the number of subjects and the number of observations in each subject tend to infinity~(\cite{Vonesh92}), which in practice is seldom verified. Alternatively, bootstrap~(\cite{Efron79}), log-likelihood profiling and Sampling Importance Resampling (SIR)~(\cite{Dosne17}) have been proposed. 

The bootstrap was initially proposed as a widely applicable jackknife approach to estimate biases and SE~(\cite{Efron79}) but then took on the more ambitious goal of building non-parametric confidence intervals~(\cite{Efron00}). It consists in repeatedly resampling the observed data with replacement and applying the original analysis to create the distribution of an estimator or a statistic of interest. The simplest method is the Case bootstrap where the entire vector of individuals is resampled. Given the hierarchical nature of the NLMEM, different approaches have been advocated to ensure a proper treatment of the inter-individual and within individual variabilities (\cite{Das99}). 
\cite{Thai13, Thai14} investigated different bootstrap approaches including the parametric bootstrap (Par), resampling residuals from a theoretical distribution, and the non-parametric bootstrap (NP) resampling from the estimated residuals based on the Empirical Bayesian Estimates (EBE) of the parameters. They found that all the bootstraps performed better than the asymptotic method in NLMEM, none showing a clear advantage. One issue with the NP bootstrap however is shrinkage due to the regression to the mean of individual parameters estimates towards population parameters~(\cite{Karlsson07}), which led to degraded performances especially in very non-linear models, despite correcting the variance of the residuals to account for shrinkage~(\cite{Carpenter03}). To overcome this, \cite{Comets21} proposed a residual non-parametric bootstrap using samples from the conditional distribution of the individual parameters~(\cite{Lavielle16}) to resample the random effects. The objective of the present study is to extend this preliminary implementation to a full conditional non-parametric bootstrap (cNP) also resampling the residual errors, since these are affected by $\epsilon$-shrinkage~(\cite{Karlsson07}).

For comparability, we will continue to use the simulation setting in~\cite{Comets21}, inspired by~\cite{Plan12}. In a first step, we implement the full cNP and compare the performances with respect to the Case, NP and Par bootstraps using the previous simulated data in rich (4 doses) and sparse (2 doses) scenarios. In a second step, we challenged the bootstraps with unbalanced designs where the amount of information differs across subjects. In a third step, we evaluate the impact of increasing the residual variability.

\section{Materials and methods} \label{sec:2.Methods}

\subsection{Statistical model} \label{sec:2.1Models}

\hskip 18pt We consider non-linear mixed effect models~(\cite{LavielleMonolix}). Let the random variable $y_{ij}$ $(i=1, \dots, N, j=1, \dots, n_i)$ denote the $j^{\rm th}$ observation in the $i^{\rm th}$ subject, $\xi_{ij}$ the corresponding design variables (eg measurement times and dose amounts in a PK study) and $z_i$ individual covariates. For models describing the evolution of continuous data, we generally assume that the observations $y_{ij}$ follow a normal distribution with a mean defined by a structural model $f$ denoting the longitudinal evolution and depending on the individual parameters $\psi_i$, and a standard deviation defined by an error model $g$ which involves residual error coefficients denoted by $\sigma$:
\begin{equation} \label{eq:normal}
y_{ij} =f(\xi_{ij}, \psi_i) + g(\xi_{ij},\psi_i, \sigma)\cdot\epsilon_{ij}
\end{equation}
where the residual errors $\epsilon_{ij}$ are i.i.d normally distributed random variables. Individual parameters $\psi_i$ are assumed to follow a multivariate distribution with known shape and unknown parameters, through a transformation $h$ of a linear function of the fixed effects $\mu$, possibly covariate effects $\beta$ and random effects $\eta_i$, which are assumed to be normally distributed with mean 0 and variance-covariance matrix $\Omega$:
\begin{equation} \label{eq:hierarchical_model}
\psi_i = h(\mu, \beta, z_i, \eta_i) \text{ where } \eta_i \sim N(0, \Omega)
\end{equation}
Standard models for $g$ include $g=\sigma$ (constant variance), $g(\xi_{ij},\psi_i, \sigma)=\sigma \: f(\xi_{ij},\psi_i)$ (proportional variance), or a combination of both.

In a frequentist framework, which will be our setting here, we estimate the population parameters $\theta = (\mu, \beta, \Omega, \sigma)$. Here, we use the iterative SAEM algorithm (\cite{Kuhn05,LavielleMonolix}), where during the stochastic approximation step the unknown parameters $\eta_i$ representing the random effects are sampled from their conditional distributions to create a complete dataset from which the conditional log-likelihood is derived by stochastic approximation. At the end of the estimation process, individual parameters are estimated as the mode or the median from the conditional distribution obtained, refined at the end of the fit by repeated sampling through an Metropolis-Hastings procedure.

Parameter uncertainty is usually assessed in the software performing maximum likelihood estimation by evaluating the Fisher Information Matrix at the point estimates $\hat{\theta}$ and using the inverse as the variance-covariance matrix yielding the standard errors. Various approximations have been proposed to ease the computation~(\cite{Retout02, Nyberg15}), the most commonly used being a first-order approximation yielding a block structure with one block for the fixed effects and one block for the parameters associated with the variabilities. A normal approximation then gives standard $\alpha$-level confidence intervals (CI) for the $p^{\rm th}$ component of $\theta$, denoted $\theta^{(p)}$, as:
\begin{equation} \label{eq:normalCI}
\theta^{(p)} \in \left[ \hat{\theta}^{(p)} - z_{\alpha/2} \hat{SE}(\hat{\theta}^{(p)}); \hat{\theta}^{(p)} + z_{1-\alpha/2} \hat{SE}(\hat{\theta}^{(p)});  \right]
\end{equation}
where $z_q$ is the $q$-quantile of the normal distribution. In the following, we will denote this approach as the Asymptotic method and drop the subscript $(p)$ for easier reading.

\subsection{Bootstrap methods}\label{sec:2.2Bootstrap}

\hskip 18pt Bootstrapping generates multiple pseudo-samples mimicking the distribution of the original sample. Denoting $B$ the number of bootstrap samples, a general bootstrap algorithm is:

\begin{enumerate}
  \item Generate a bootstrap sample by resampling from the data and/or from the estimated model
  \item Compute the estimates of the parameters of the model for the bootstrap sample
  \item Repeat steps 1-2 $B$ times to obtain the bootstrap distribution of the parameter estimates.
\end{enumerate}
The mean and standard deviation of the distribution of the bootstrap estimates $\hat{\theta}_b^{*}$ ($b=1...B$) can be viewed as the bootstrap estimates of the parameters and their SE. These could be used to derive a CI through the normal approximation as in equation~\ref{eq:normalCI}, however the bootstrap distribution may be asymmetric contrary to the asymptotic estimate and it is more appropriate to compute the non-parametric or percentile CI, defined as the bootstrap $\alpha$-level CI from the bootstrap distribution~(\cite{Davison97}):
\begin{equation}
\label{eq:Boot_CI}
\theta \in \left[ \hat{\theta}^{*}_{\frac{\alpha}{2}}  ;\hat{\theta}^{*}_{(1-\frac{\alpha}{2})} \right]
\end{equation}
where $\hat{\theta}^{*}_{(\frac{\alpha}{2} \cdot B)}$ is the $\frac{\alpha}{2}$-quantile of the bootstrap distribution. Other approaches to construct CI, including Student or normal-type approximations as well as bias-corrected CI, are discussed in particular in~\cite{Davison97} and~\cite{McKinnon06}.

\bigskip
Different approaches have been proposed to generate bootstrap samples in NLMEM to reflect the true data generating process with repeated measures within a subject and take into account the two levels of variability. The case bootstrap (Case) resamples entire observation vectors by resampling individuals. Parametric (Par) and non-parametric (NP) bootstraps on the other hand resample random effects and residual errors to reconstruct an observation vector, preserving the structure of the original dataset (detailed algorithms are given in online Appendix~\ref{App_bootstrapMethods}). Here, we extend the partial conditional non-parametric residual bootstrap using samples from the conditional distribution~(\cite{Comets21}) to also resample the residual errors, yielding the following algorithm (cNP):
\begin{enumerate} 
  \item fit the model to the data to obtain $\hat{\theta}$ and estimate the conditional distributions
  \item obtain $M$ samples for each vector $\eta_i$, $\eta_{i}^m$ ($m=1,...,M$) from the conditional distributions
  \item calculate the residuals $\epsilon^{m}_{ij} = \frac{y_{ij} - f(x_{ij}, \hat{\mu}, \eta^{m}_{i}) }{g(x_{ij}, \hat{\mu}, \eta^{m}_{i},\sigma)}$, then center them
  \item center the $\eta_{i}^m$ for each subject and draw a sample $\{ \eta^*_i \}_{i=1,...N}$ with replacement 
  \item draw a sample \{$\epsilon^{*}_{ij}$\} with replacement globally from \{$\epsilon^{*m}_{ij}$\} 
  \item generate the bootstrap responses $y^{*}_{ij}=  f(x_{ij}, \hat{\mu}, \eta^*_i) + g(x_{ij}, \hat{\mu}, \eta^*_i, \sigma) \; \epsilon^*_{ij}$
\end{enumerate}
In the present study we performed the resampling of residuals in steps (b) and (d) separately for each subject within their own distribution, but in homogeneous populations it would also be possible to sample globally over the residuals obtained for the entire population.

\subsection{Simulation study} \label{sec:2.3Simulation}

We performed three sets of simulations to evaluate our new bootstrap approach. First, we applied the cNP to the previous simulations used in the assessment of the partial cNP for random effects in~\cite{Comets21}. Second, we investigated the influence of the amount of information by simulating unbalanced designs. Third, we investigated the effect of increasing the residual variability.

\paragraph{Simulation model:} we used the model initially proposed by \cite{Plan12} in a comparison of several algorithms, where different scenarios allowed to investigate the influence of non-linearity and design on the performance of estimation methods. The data follows a dose-response model:

\begin{equation} \label{eq:continuous_structural_model}
f(x_{ij},\psi_i)={\rm E}_{0_i} +{\rm E}_{max_i} \cdot \frac{x_{ij}^\gamma}{x_{ij}^\gamma+{\rm ED}_{50_i}^\gamma}
\end{equation}

This model involves 4 parameters, the initial effect E$_0$, the maximum effect E$_{\rm max}$, the concentration at which half the maximum effect is achieved ED$_{50}$ and the sigmoidicity factor $\gamma$ which controls the non-linearity of the model through the curvature. The parameters were assumed to follow a log-normal distribution, except for $\gamma$, which was assumed to be the same for all subjects (no variability), so that $\psi_i=\{ {\rm E}_{0,i}, {\rm E}_{max,i}, {\rm ED}_{50,i}, \gamma \}$. A covariance was simulated between E$_{\rm max}$ and ED$_{50}$, and a proportional error model with coefficient $\sigma$ was used. Two values of $\gamma$ were used, $\gamma=1$ (E$_{\rm max}$ model) and $\gamma=3$ (Hill model) to evaluate how model non-linearity affected the results. The values of the parameters used in the simulations are presented in Table \ref{tab:continuous_parameters}.

\paragraph{Simulation scenarios:} In the first scenario, we used the datasets already simulated for the paper by \cite{Comets17} (available online in a \href{https://zenodo.org/record/4059718#.YM0302gzZPY}{\color{blue}Zenodo repository}), which included a rich (4 doses) and sparse (2 doses) design for each of the two models (E$_{\rm max}$ with $\gamma=1$ and Hill with $\gamma=3$). In the second scenario, we considered unbalanced designs representing situations where some subjects receive a full set of doses while others receive only low, only high, or a mix of doses, as well as a mix of sparse (2 doses) and rich (4 doses) design, as illustrated in Figure \ref{fig:unbalanced}. The third scenario used the same settings as the first scenario but increasing $\sigma$ to 0.3 and 0.5. Detailed description of the designs are given in online Appendix~\ref{App_simulation}. For each dataset and method, $B=200$ bootstrap samples were used. This number could be increased to better approach extreme quantiles~(\cite{Austin20}). 

\paragraph{Bootstrap evaluation:}

\hskip 18pt For each simulated dataset $k=1, 2, ..., K$ we computed the following statistics of interest obtained from $B$ bootstrap samples by the four methods, for each estimated parameter: the bootstrap mean $\hat{\theta}_{B,k}$, the bootstrap standard error $\widehat{SE}(\theta)_{B,k}$, obtained as the standard deviation of the bootstrap distribution $\widehat{SE}(\theta)_{B,k}=SD(\hat{\theta}_{B,k})$, and the bootstrap CI obtained using the quantiles of the bootstrap distribution. 

Our main evaluation metric was the coverage rate for the 90\% CI obtained by the four bootstraps and the Asymptotic method, calculated using the SE estimated by {\sf saemix}. The coverage rate of a CI of parameter is the proportion of times that it contains the true value of the parameter, reflecting the accuracy of each approach to estimate uncertainty. The 90\% CI ($\alpha=0.1$) was chosen to allow visualising both undercoverage and overcoverage. We also report the results for the more standard 95\% CI ($\alpha=0.05$) in online Appendix~\ref{App_originalCR}. Fluctuations due to the simulations was assessed by computing the SE related to the Monte Carlo procedure as suggested by \cite{Morris19}, and used to choose the number of simulations ($K$). 
\begin{equation} \label{eq:n_sim}
MC.SE=\sqrt{\frac{E(coverage)(1-E(coverage))}{K}}
\end{equation}
We chose $K=200$ which yielded an expected MC.SE of approximately 2\%.

\subsection{Implementation} \label{sec:2.4 Implementation}

\hskip 18pt Simulations were performed in R (\cite{R}) using R version 4.0.4(2021-02-15). The parameters of the models were estimated using the {\sf saemix} library (\cite{Comets17}) in the development version available on its \href{https://github.com/saemixdevelopment/saemixextension}{\color{blue}github repository}. We implemented the new bootstrap approach in the development version of {\sf saemix}. The default options were used to run the SAEM algorithm; the initial parameters for the fixed and random effects were set to the true value in all runs as the objective was not to evaluate the estimation method but the bootstrap performances. The results were analysed in R. The runs were set up using a laptop and the simulation studies were run on the \textit{"Centre de Biomodélisation IAME"} (CATIBioMed cluster, Bichat hospital, Paris).  For the evaluation of the designs we used the R package PFIM 4.0 (\cite{Dumont18}) available on the \href{http://www.pfim.biostat.fr/}{\textcolor{blue}{PFIM website.}} where we evaluated the scenarios presented in this study, minus the covariance term as the software does not allow a full covariance matrix. The Zenodo repositories containing the simulated datasets used in this study have DOI 10.5281/zenodo.4059718 and 10.5281/zenodo.10779192.

\section{Results} \label{sec:3Results}

\subsection{Original designs} \label{sec:3.1Original}

\hskip 18pt In Figure \ref{fig:CR90_original} we plot the coverage rates of 90\% CI (points) along with their MC uncertainty (errorbars) for the four designs in the original simulation scenario. Non-parametric CI are presented with a different colour for each method. We observe similar patterns as in the previous analysis for the 95\% CI~(\cite{Comets21}), with the asymptotic method showing significant undercoverage for several parameters. The sparse design with a Hill model ($S_{Hill, S}$) was the most challenging case, with all bootstraps failing to get good coverages for at least one parameter. In this scenario the case bootstrap showed the most overcoverage, which wasn't apparent previously with a 95\% CI. 

Our proposed cNP performs better than the classical NP bootstrap across the scenarios and as well as the Case and the Parametric bootstrap, although the CI for the fixed effects are larger in S$_{\rm Hill,S}$ scenario. The NP underestimates the coverage for some parameters in every design, especially in $S_{Hill,R}$. Online appendix~\ref{fig:Bias_original} shows plots of the bias with respect to the true values for the different approaches (left column), and we can see that for the two rich designs the estimation by {\sf saemix} is unbiased for all parameters, suggesting that the poor performance with NP is not due to biased estimates of the population parameters. The plots in the right hand column in online Appendix~\ref{fig:Bias_original}, which compare the SD of the bootstrap distributions to the empirical SE, show that the SE tend to be estimated with a bias around $\pm$10\% (depending on the parameter) which could indicate that the NP has more difficulties in narrowing on the extreme quantiles,  while the cNP may be able to correct for this using samples accepted by the Metropolis-Hastings algorithm. The same plots provide a possible explanation for the undercoverage observed for the Asymptotic method, as they show that this method consistently underestimates the SE for most designs.


\subsection{Unbalanced designs} \label{sec:3.2Unbalanced}

\hskip 18pt Figure \ref{fig:CR90_unbalanced_Hill} shows the same coverage rate with the Hill model for the four unbalanced designs (see online Appendix~\ref{App_unbalanced_CR} for the E$_{\rm max}$ scenarios, and the 95\% coverage rates under both models).

Across all unbalanced designs and as previously, the asymptotic method showed some undercoverage, but less important probably due to the designs being more informative (online Appendix~\ref{App_unbalanced_designeval}. The bootstraps generally had adequate coverage, except for the NP for the design $S_{Hill,end}$. The Case bootstrap in this case was not stratified as the results were already good without stratification. To investigate further we plotted the relative bias on the parameters and their SE compared to the true values (figure \ref{fig:Bias_unbalanced_Hill} in online Appendix \ref{App_unbalanced_Bias}) and found that for this design, both the Par and NP bootstrap showed significant bias for the random effects both for the parameter estimate and for the SE, while the Case and the cNP bootstraps were within $\pm10\%$. This trend was apparent across all designs, with Par and NP presenting more bias in the unbalanced designs compared to the original designs. 

\subsection{Increased error} \label{sec:3.3Error}

\begin{figure}[!ht]
    \centering
    \makebox{\includegraphics[scale=0.9]{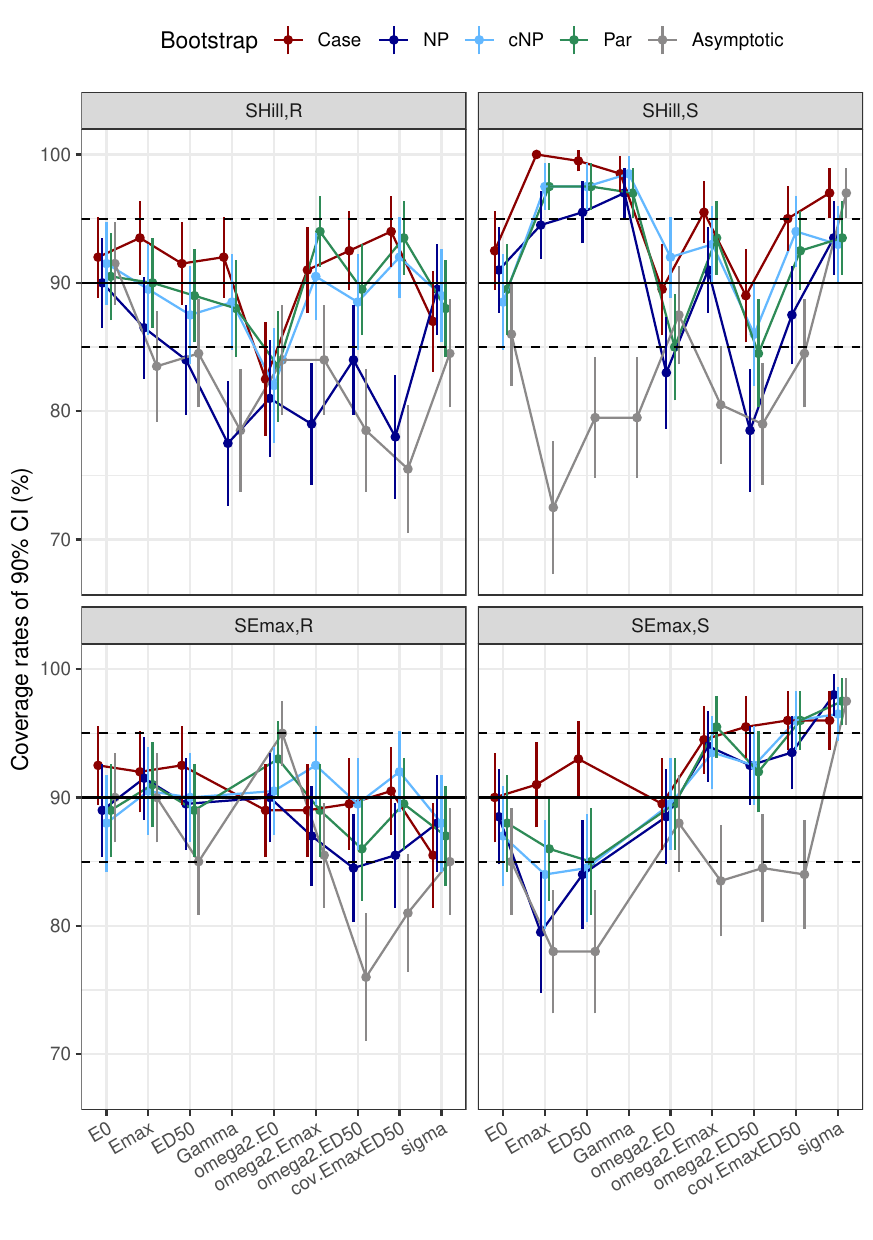}}
    \caption{\label{fig:CR90_original}Coverage rates of the 90\% CI along with the errorbars of their MC uncertainty for the original designs. Dashed lines indicate a coverage of 85 and 95\%. The same scale is used on the Y-axis for the 4 graphs. The X-axis is jittered to avoid superimposing the different approaches.}
\end{figure}

\begin{figure}[!h]
    \centering
    \makebox{\includegraphics[scale=0.85]{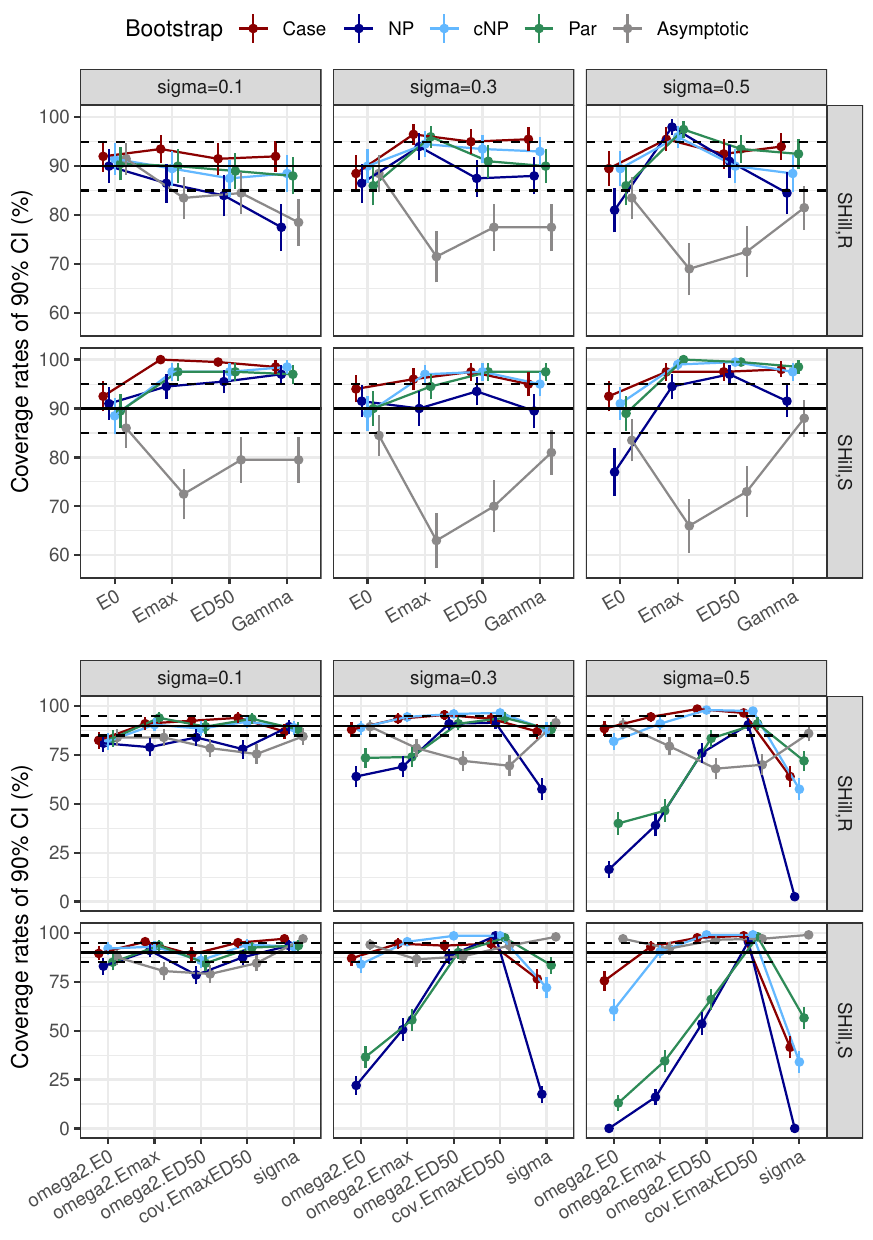}}
    \caption{\label{fig:CR90_error_Hill}Coverage rates of the 90\% CI and errorbars of their MC uncertainty for the original design and the two designs with increased error coefficient $\sigma=0.3$ and $\sigma=0.5$ under the $S_{Hill}$. Dashed lines indicate a coverage of 85 and 95\%. The same scale is used on the Y-axis for the 4 graphs. The X-axis is jittered to avoid superimposing the different approaches. At the first pair of lines we show the fixed effects and at the second the random effects. Each pair of lines has the same scale on the Y-axis but we separated the fixed effects (top two lines) and the variance parameters (bottom two lines) because of the different scales.}
\end{figure}

\hskip 18pt Figure \ref{fig:CR90_error_Hill} contrasts the coverage rates for the 90\% CI in the rich and sparse Hill designs (see online Appendix~\ref{App_error_CR} for additional results) from the original scenario ($\sigma=0.1$, left) with the scenarios with increased error coefficient $\sigma=0.3$ and $\sigma=0.5$ (middle and right columns). The undercoverage already observed in figure~\ref{fig:CR90_original} for the coverage rate of fixed effects by the Asymptotic method becomes more pronounced as $\sigma$ increases, affecting all parameters except E$_0$. Online Appendix~\ref{App_error_Bias} shows that this may again be linked to underestimation of the SE, and is also logically higher in the sparse design. The behaviour of the NP boostrap is somewhat inconsistent, sometimes correcting the poor coverages observed with $\sigma=0.1$ and sometimes inflating them. 

For the variance parameters, in the rich design the asymptotic method again underestimates the SE leading to undercoverage, but this time we see a marked degradation of both the NP and more surprisingly of the Par bootstrap as $\sigma$ increases. The Case and the cNP perform generally better but show undercoverage for $\omega_{{\rm E}_0}$ and $\sigma$. At the highest level of residual error tested here all bootstraps provided very poor coverage for $\sigma$, with a largely underestimated uncertainty for this parameter. The same pattern can be seen for the sparse design, but there the asymptotic method only shows a slight overestimation for the random effects with less impact on the SE, and as a consequence the coverage rate are adequate. The plots of bias compared to the true parameters in online Appendix~\ref{App_error_Bias} in both the rich and sparse designs show that both the NP and the Par bootstraps tend to amplify the bias from the original estimates, which are used to calibrate the theoretical distributions for Par and as priors for NP, and can explain why these bootstraps behave poorly, while both the Case and cNP rely more strongly on the data and can partly correct for the bias. 

\section{Discussion} \label{sec:Discussion}

\hskip 18pt The aim of this study was to extend the conditional non-parametric bootstrap approach for NLMEM proposed in \cite{Comets21} by also resampling the residual error, and evaluate this full cNP by comparison to three bootstrap methods previously used in NLMEM~(\cite{Thai14}), case (also called pairs bootstrap), non-parametric and parametric bootstraps. We also investigated the influence of an imbalance in the number of samples across subjects and of the magnitude of the residual error. Coverage rates were our main evaluation metric since the bootstrap is a non-parametric approach, and the bootstrap distributions may deviate from the Gaussian distribution obtained through a normal approximation. Additional results are given in online Appendices for the bias on parameters using the mean value of the bootstrap distribution as the bootstrap estimates, and the comparison between the empirical SE and the standard deviation of the bootstrap distribution which is often used as an approximation to the SE. 

The cNP uses the conditional distributions, obtained as a by-product of the estimation process in the SAEM algorithms~(\cite{Lavielle16}). Samples from the conditional distribution are used to produce informative diagnostics by avoiding shrinkage when the number of individual samples is small. Here the comparison between cNP and NP shows that they allow to generate more representative bootstrap datasets than using empirical individual residuals with a population-level correction~(\cite{Carpenter03, 
Morris02}). 
This could mean that the exchangeability assumption, a core bootstrap principle, is better respected~(\cite{Davison97}), as the cNP resamples residuals within their own conditional distributions which have equal sampling probability. 

A major motivating factor behind the development of the cNP was its ability to generate bootstrap samples with the same design as the original dataset, since resampling at the level of the residuals preserves the original design and covariate structure without needing to stratify over variables of interest. Case bootstrap, widely used for its simplicity and general applicability, requires adjusting test statistics and does not preserve the design~(\cite{McKinnon06}). Here, when we simulated unbalanced designs with a varying amount of information per subject, both cNP and Case estimated the parameters with low bias, except in S$_{\rm Hill,low}$ for the covariance term, where it was limited enough to avoid affecting the coverage. In general, all bootstraps performed well in the unbalanced designs, and the Case bootstrap did not need a stratification to maintain appropriate coverages. 
The predicted RSE (online Appendix~\ref{App_unbalanced_Bias}), which we used during the design of the simulation study to ensure parameters were identifiable, suggest that the unbalanced designs were probably in fact not challenging enough, especially when we compare to the predicted RSE from the original design. Stratification could be explored in scenarios involving covariates in the model, but our initial results suggest that the Case bootstrap remains robust.

All bootstraps however failed to some degrees in the final simulation setting with high $\sigma$. The plots of bias shown in the online Appendices show an increased bias for the variance parameters, most affected by the undercoverage, indicating that the bootstrap distributions move away from the true value. More generally, bootstrap methods find it difficult to recover from poor estimates on the original datasets, and we even found that Par and NP amplify the bias in the asymptotic method (see for example fig~\ref{fig:CR95_error_Emax} and~\ref{fig:CR95_error_Hill} in online Appendix~\ref{App_error}). This corroborates other works showing that bootstrap approaches don't perform well in small samples, where there is not enough information to estimate variabilities: in~\cite{Broeker20}, the Case bootstrap was implemented and performed very poorly in small samples, especially compared to an approach combining univariate log-likelihood profiling and Sampling Importance Resampling (\cite{Dosne17}) or a full Bayesian approach. In the present paper with larger sample sizes, the Case and cNP generally performed similarly and were least sensitive in this scenario, so that it would be interesting to compare the cNP in the setting of~\cite{Broeker20}. The poor performance of Par was somewhat unexpected, as it samples in a distribution with the true shape, but it indicate a high sensitivity to biased estimates from the estimation on the original data, even amplifying it as the bootstrap estimates drift further away from the asymptotic estimates. Because the cNP uses an acceptation-rejection algorithm to sample from the conditional distribution, it may be able to correct better for the drift. One final note is that we did not examine each bootstrap distribution produced in this simulation study, but in practice, we would recommend to plot the bootstrap distributions as their shape and support space can be an indication of the stability from one bootstrap run to another. Patently aberrant estimates can occur especially in sparse designs with high variance, and could be excluded before building the confidence intervals. Another necessary check is model adequacy which can be examined by diagnostics~(\cite{Nguyen17}) as well as run assessment~(\cite{LavielleMonolix}) to check the sensitivity of the model estimates to different initial starting values and random seeds. 


\section{Conclusion} \label{sec:Conclusion}

\hskip 18pt We proposed a new bootstrap approach for non-linear mixed effect models using samples from the conditional distribution instead of estimated residuals. We implemented it in version 3.2 the {\sf saemix} package for R, and compared its performances to other widely used bootstrap methods, for a model with continuous outcomes. The cNP bootstrap improved the coverage rate in the different scenarios tested in this study over the traditional non parametric bootstrap, and was as efficient as the more simple case bootstrap. Although it is more computationally intensive than Case bootstrap, it has the advantage of preserving the structure of the original dataset, including the covariate distribution and design, which make it conceptually simpler to implement in complex models where Case bootstrap may need stratification to maintain identifiability.

Bootstrap approaches however are not a magic wand and cannot compensate for poor estimates, such as variance terms in small sample size or poorly identifiable parameters. In particular we recommend performing thorough diagnostics of model adequacy and stability before applying a bootstrap approach.


\section*{Acknowledgments}

The authors would like to thank Hervé Le Nagard, Lionel de la Tribouille and Rémy Bertino for the use of the CATIBioMed computer facility.

\section*{Disclosure Statement}

\noindent{\bf Author contributions:} E.C. and S.K. designed the research. S.K. performed the research, analysed the data and wrote the first draft of the manuscript. M.U. and E.C. discussed and interpreted the results. E.C. finalised the manuscript with M.U. and S.K. providing input and discussion elements.

\noindent{\bf Funding:} No funding was received for this work. E.C. has received consulting fees from Sanofi, bearing no relationship with the present manuscript.

\renewcommand\bibname{References}
\renewcommand\refname{References}
\bibliography{references}
\bibliographystyle{apalike}

\newpage

\section*{Tables}
\begin{table}[!ht]
    \caption{\label{tab:continuous_parameters} Parameters used in simulations}
    \begin{center}
    \begin{tabular}{l c c c c}   
    \toprule
    {\bf Parameter} & {\bf Value} && {\bf Parameter} & {\bf Value} \\
        \hline
         E$_0$ & 5 && $\omega_{{\rm E}_0}^2$ & 0.09 \\
         E$_{\rm max}$ & 30 && $\omega_{{\rm E}_{max}}^2$ & 0.49\\
         ED$_{50}$ & 500 && $\omega_{{\rm ED}_{50}}^2$ & 0.49 \\
         $\gamma$ & 1 or 3 && ${cov}({\rm E}_{max},{\rm ED}_{50})$ & 0.245\\
         &&& $\sigma$ & 0.1 \\
        \bottomrule
     \end{tabular}
         \end{center}

\end{table}

\section*{Figures}

\begin{figure}[!ht]
    \centering
    \makebox{\includegraphics[scale=0.99]{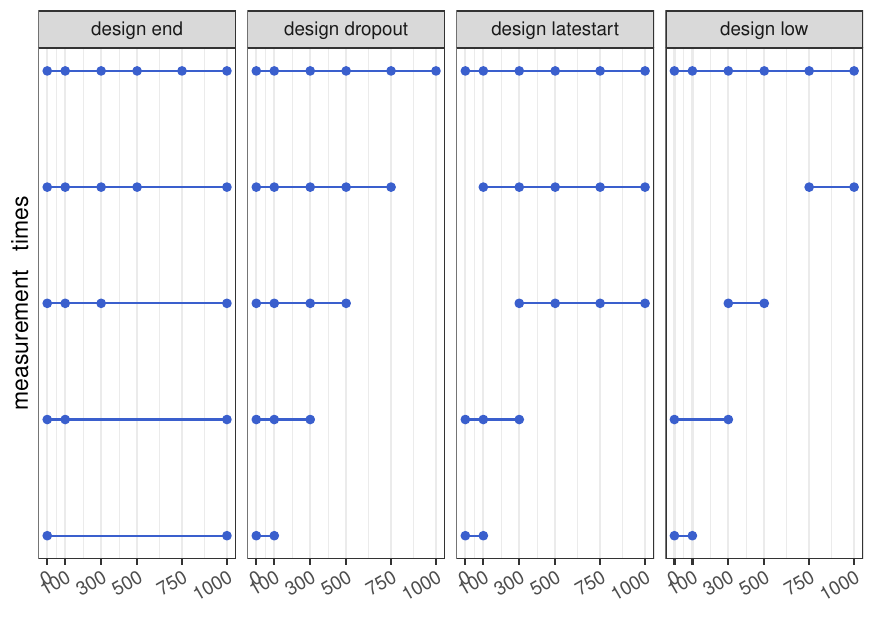}}
    \caption{\label{fig:unbalanced}: Measurement times in the four unbalanced designs. The first 3 designs involved 100 subjects divided in 5 groups of 20 with 2 to 6 doses, and the last group included 80 subjects with 2 doses and 20 subjects with 6 doses.}
\end{figure}

\newpage

\newpage
\begin{figure}[!h]
    \centering
    \makebox{\includegraphics[scale=0.9]{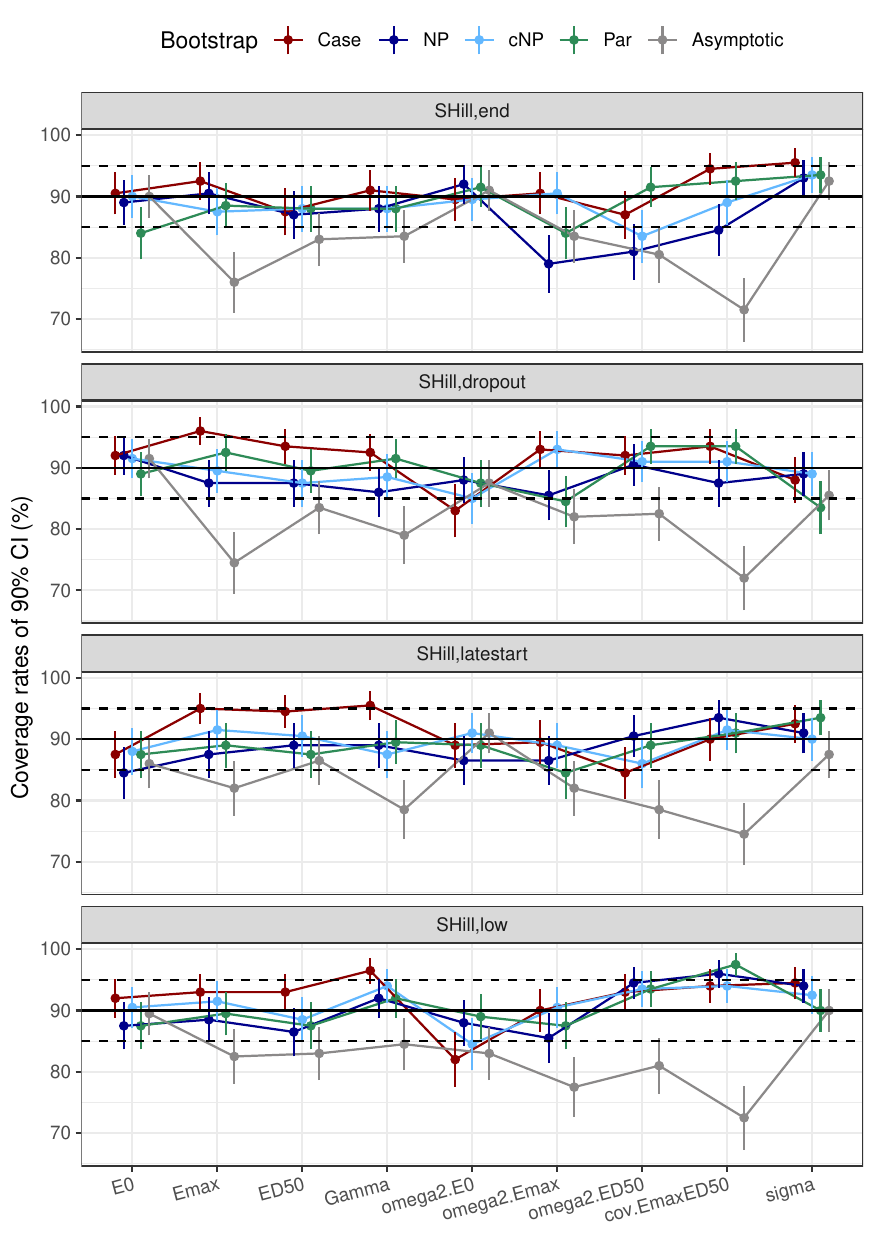}}
    \caption{\label{fig:CR90_unbalanced_Hill} Coverage rates of the 90\% CI and errorbars of their MC uncertainty for the unbalanced designs for the $S_{Hill}$. Dashed lines indicate a coverage of 85 and 95\%. The same scale is used on the Y-axis for the 4 graphs. The X-axis is jittered to avoid superimposing the different approaches.}
\end{figure}

\newpage

\clearpage
\newpage
\appendix
    \begin{center}
  {\bfseries \Large Full conditional non-parametric bootstrap}
  \\ 
    {\bfseries \Large   an evaluation with unbalanced designs and high residual variability}
    \\
  \vspace{1cm}
  {\bfseries \Large Supplementary material}
  \end{center}
  
  \vspace{1cm}
    {\large 
   {\bfseries Authors:} Sofia Kaisaridi, Moreno Ursino, Emmanuelle Comets} 

\vspace{2cm}
\section{Appendix: Bootstrap algorithms} \label{App_bootstrapMethods}

\subsection*{Methods}

Bootstrapping generates multiple pseudo-samples mimicking the distribution of the original sample. Denoting $B$ the number of bootstrap samples, a general bootstrap algorithm is:

\begin{enumerate}
  \item Generate a bootstrap sample by resampling from the data and/or from the estimated model
  \item Compute the estimates of the parameters of the model for the bootstrap sample
  \item Repeat steps 1-2 $B$ times to obtain the bootstrap distribution of the parameter estimates.
\end{enumerate}

\paragraph{Case bootstrap (Case):} This method consists of resampling with replacement the entire subjects ($\xi_i$,$\textbf{y}_i$ where $\textbf{y}_i= (y_{i1},y_{i2},....,y_{in_i})'$) from the original data before modelling. It is also called the \emph{paired bootstrap}. It is the most straightforward way to do bootstrapping and is mostly assumption-free. Stratification may be applied when the data is non-homogeneous to preserve the distribution of a covariate or of a sampling schedule.

\paragraph{Parametric residual bootstrap (Par):} The parametric bootstrap requires the strongest assumptions as it depends both on the model and the distributions of parameters and errors. This method resamples the residuals by simulating from the theoretical distributions obtained after fitting the model, here for the random effects from the normal distributions adjusted the estimated mean and variance of the initial fit. In non-linear mixed effect models, two levels of variability must be considered, within-subject and between different subjects. The bootstrap sample is obtained as follows:

\begin{enumerate} 
  \item fit the model to the data to obtain $\hat{\theta}$
  \item draw a sample $\{\eta^{*}_i\}_{i=1, \dots, N}$ from a normal distribution with mean zero and covariance matrix $\hat{\Omega}$
  \item draw a sample $\{\epsilon^{*}_{ij}\}_{j=1, \dots n_i}$ from a normal distribution with mean zero and covariance matrix $\mathcal{I}_{n_i}$ where $\mathcal{I}$ denotes the identity matrix
  \item generate the bootstrap responses $y^{*}_{ij}=  f(x_{ij}, \hat{\mu}, \eta^*_i) + g(x_{ij}, \hat{\mu}, \eta^*_i, \sigma) \; \epsilon^*_{ij}$ (where we explicitly replace $\psi_i$ by $\hat{\mu}$ and $\eta^*_i$ to highlight the dependency on the resampled random effects)
\end{enumerate}

\paragraph{Non-parametric residual bootstrap (NP):} This method resamples with replacement from the residuals obtained after model fitting for all subjects. The bootstrap sample is obtained as follows:

\begin{enumerate}
  \item fit the model to the data to obtain $\hat{\theta}$
  \item estimate the individual parameters $\hat{\eta}_i$ as the EBE, then center and normalise them as detailed below
  \item draw a sample $\eta^*_i$ with replacement for $i=1,...N$ from the resulting set
  \item calculate the residuals $\hat{\epsilon}_{ij}= y_{ij}-f(x_{ij}, \hat{\mu}, \hat{\eta}_i) $, then center and normalise them
  \item draw a sample $\{\epsilon^{*}_{ij}\}_{j=1, \dots n_i}$ with replacement globally from the resulting set
  \item generate the bootstrap responses $y^{*}_{ij}=  f(x_{ij}, \hat{\mu}, \eta^*_i) + g(x_{ij}, \hat{\mu}, \eta^*_i, \sigma) \; \epsilon^*_{ij}$
\end{enumerate}

The normalisation in step (b) and (d) is necessary because the EBE generally suffer from regression to the mean and their variance may be considerably smaller than the true variability in the population~(\cite{Karlsson07}). To correct for this, \cite{Carpenter03} suggested centering the residuals, to ensure their distribution has mean 0, and inflating the variance by using the ratio between the estimated and empirical variance-covariance matrices. The transformation of random effects was carried out using the eigenvalue decomposition (EVD) proposed by \cite{Thai14} to limit numerical difficulties due to almost singular matrices. This procedure reads as follows:

\begin{enumerate}
  \item center the raw estimated random effects: $\tilde{\eta_i}=\hat{\eta}_i-\bar{\eta}_i$
  \item obtain the EVD of the estimated variance-covariance matrix: $\hat{\Omega} = V_{\Omega} \; D_{\Omega} V_{\Omega}^T$ where $ D_{\Omega} $ is the diagonal matrix containing the eigenvalues of $\hat{\Omega}$
  \item obtain the EVD of $S$, the variance-covariance matrix of the centered random effects:  $S = V_{S} \; D_{S} V_{S}^T$ where $D_S$ is the diagonal matrix containing the eigenvalues of $S$
  \item calculate the correction matrix $A_{\eta}$ using these two decompositions as \\
  $A_{\eta}= V_S \; D_S^{-1/2} \; V_{\Omega} \; D_{\Omega}^{-1/2} $
  \item transform the centered random effects using the ratio $A_\eta$: $\hat{\eta}^{'}_i=\tilde{\eta}_i \times A_\eta$
\end{enumerate}

Similarly, the residuals are transformed using the empirical variance:
\begin{enumerate}
  \item center the raw estimated residuals: $\tilde{\epsilon}_{ij}=\hat{\epsilon}_{ij}-\bar{\epsilon}_{ij}$
  \item calculate the correction factor $A_{\sigma} = 1/\sigma_{emp}$ where $\sigma_{emp}$ is the empirical standard deviation of the centered residuals
  \item transform the centered residuals using the ratio $A_\sigma$: $\hat{\epsilon}_{ij}^{'}=\tilde{\epsilon}_{ij} \times A_\sigma$
\end{enumerate}

\subsection*{Implementation and runtimes}

The bootstrap methods were implemented for the saemix package in version 3.2~(\cite{Comets17}) and is available on CRAN. Development versions are downloadable on github \url{https://github.com/saemixdevelopment/saemixextension}.

In terms of runtimes, the cNP was slower than the other bootstraps by 30 to 100\% in the various scenarios, with the limiting step being the computation of the conditional distribution. However, although this made our simulations comparatively slower for this bootstrap than for the other residual-based bootstraps, in practice for a real data analysis the conditional distribution is a recommended step in the estimation as it makes the diagnostic graphs more informative, therefore the additional computational cost incurred by using cNP is mostly taken care of during the estimation itself and can be leveraged for better model diagnostics and a more precise estimation of the log-likelihood using importance sampling. The computation time increases with the number of bootstraps. Here we used 200 bootstrap samples for each approach to keep the simulations manageable and because a calibration in~\cite{Comets21} showed the coverages to be stable, but the number of samples should be increased when considering more extreme quantiles. 

\section{Appendix: Simulation scenarios} \label{App_simulation}

In the first scenario, a rich and a sparse design were simulated for each model (E$_{\rm max}$ with $\gamma=1$ and Hill with $\gamma=3$). The rich design included N=100 subjects given 4 doses each (0, 100, 300, 1000), and the sparse design N=200 subjects divided in 4 groups of 50 subjects given 2 doses each among the following 4 combinations (0,1000), (100,1000), (0,300) and (100,300). 

In the second scenario, we considered unbalanced designs with non-homogenous amount of information across subjects. We evaluated using the Fisher information matrix (FIM) several combinations from the dosing set (0, 100, 300, 500, 750, 1000), with 100 subjects divided in 5 groups of 20 with 2 to 6 doses to select the 3 designs shown in the first 3 facets of Figure \ref{fig:unbalanced}, representing situations where some subjects receive a full set of doses while others receive only low, only high, or a mix of doses. We simulated a last scenario (last facet of the figure) where 80 subjects received only 2 doses while 20 subjects received the full set of doses, to mimic a mix of sparse and rich designs. The expected RSE according to PFIM was reasonable (see online Appendix~\ref{App_unbalanced_designeval}). 

In the third scenario, we performed two additional simulations following the original designs (rich with N=100, $n_i$=4, and sparse with N=100, $n_i$=2), where we increased the error coefficient $\sigma$ to 0.3 and 0.5, leaving the other parameters unchanged.
In Table \ref{tab:simulation scenarios} we present the details for each simulation scenario in this study.

\begin{table}[!h]
    \caption{\label{tab:simulation scenarios}Simulation scenarios}
    \centering
    \fbox{%
    \begin{tabular}{c c c c c}
    {\bf Simulation} & {\bf Design} & & {\bf Subjects} & {\bf Doses}\\
        \hline
         Original & Rich && 100 & 4 \\
         & Sparse && 200 (groups of 50) & 2\\
         &&&&\\
         Unbalanced &  && 100 (groups of 20) & 2,3,4,5,6 \\
         &&&&\\
         Increased error & Rich && 100 & 4\\
         {\scriptsize ($\sigma$=0.3 or 0.5)}& Sparse && 200 (groups of 50) & 2 \\
    \end{tabular}}
\end{table}

\newpage

\section{Appendix: Original designs} \label{App_original}

\subsection{Coverage rates} \label{App_originalCR}

\begin{figure}[!ht]
    \centering
    \makebox{\includegraphics[scale=0.9]{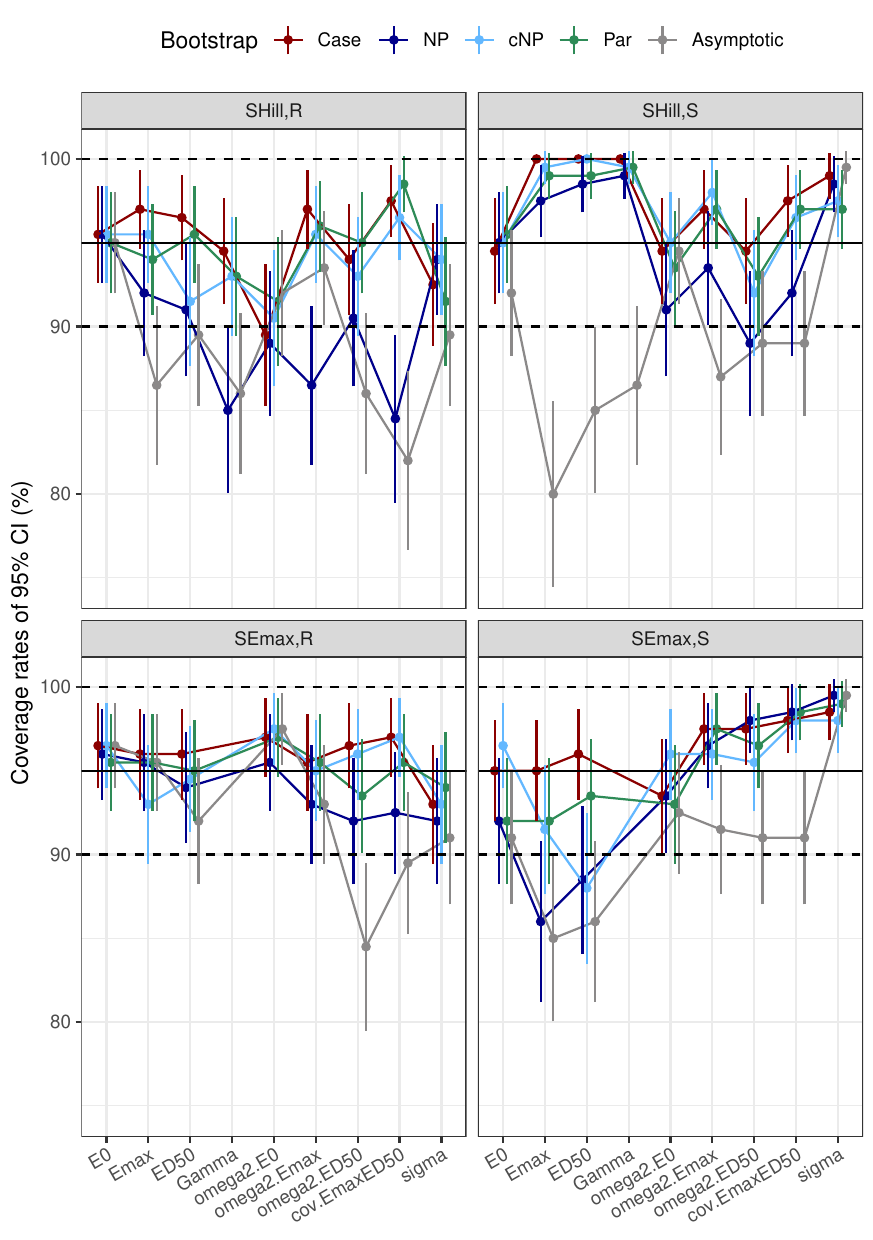}}
    \caption{\label{fig:CR95_original}Coverage rates of 95\% CI and errorbars of their MC uncertainty for the original designs. Dashed lines indicate a coverage of 90 and 100\%. The same scale is used on the Y-axis for the 4 graphs. The X-axis is jittered to avoid superimposing the different approaches.}
\end{figure}

In figure \ref{fig:CR95_original} we plot the coverage rates for the 95\% confidence intervals and their MC uncertainty for the original simulation scenarios.

\subsection{Bias} \label{App_originalBias}

\begin{figure}[!h]
    \centering
    \makebox{\includegraphics[scale=0.99]{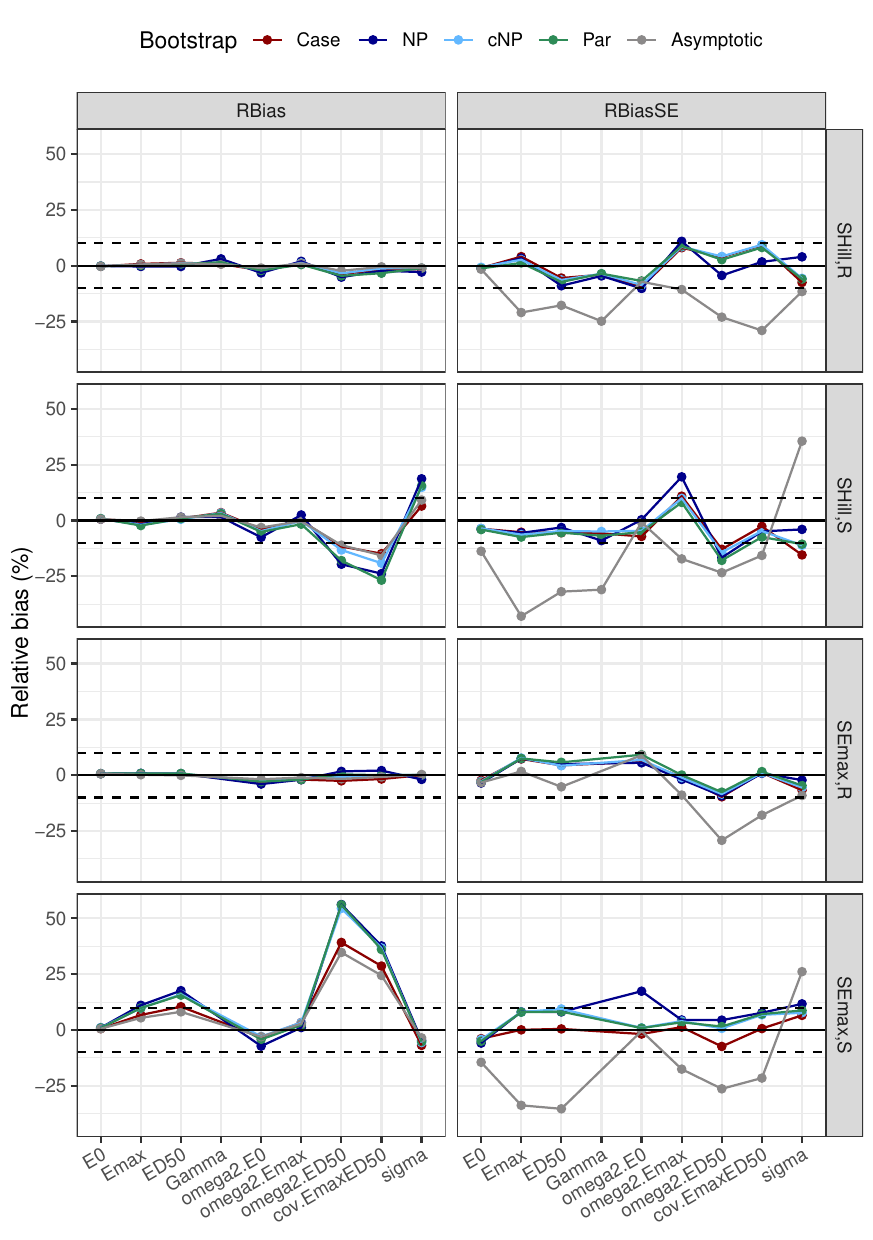}}
    \caption{\label{fig:Bias_original}Relative bias on parameters (left) and SE (right) for the original designs. Dashed lines delineate absolute relative biases within 10\%. The same scale on the Y-axis is used}
\end{figure}

Figure \ref{fig:Bias_original} shows the bias on the parameters and their S.E. obtained by all the bootstraps as well as the asymptotic method for the original designs. We calculate the bootstrap relative bias (RB) for the parameters with respect to the true values $\theta_0$:

\begin{equation} \label{eq:Bootstrap_RB_true}
RBbias(\hat{\theta})_{true}=\frac{1}{K}\cdot\sum_{k=1}^{K}\frac{\hat{\theta}_{B,k}-\theta_0}{\theta_0}\times100   
\end{equation}

We compared the bootstrap SE with the empirical SE obtained by K simulated datasets giving the "true" value observed across the simulations:

\begin{equation} \label{eq:SE_emp}
{\rm SE}_{\rm empirical}(\hat{\theta})=\sqrt{\frac{1}{K-1}\cdot\sum_{k=1}^K(\hat{\theta}_k-\theta_0)^2}    
\end{equation}

where $\theta_k$ is the value of the parameter estimated in the $k^{th}$ dataset and $\theta_0$ the true value. The relative bias on SE by each bootstrap method is then computed as the percentage difference between the average bootstrap SE and the empirical SE:

\begin{equation} \label{eq:Bootstrap_RB.SE}
RBbias({\rm SE}(\hat{\theta}))=\frac{1}{K}\cdot\sum_{k=1}^{K}\frac{\widehat{\rm SE}_{B,k}-{\rm SE}_{\rm empirical}(\hat{\theta})}{{\rm SE}_{\rm empirical}(\hat{\theta})}\times100   
\end{equation}

\newpage

\clearpage
\newpage
\section{Appendix: Unbalanced designs} \label{App_unbalanced}

\subsection{Design evaluation} \label{App_unbalanced_designeval}

\begin{figure}[!ht]
    \centering
    \makebox{\includegraphics[width=12cm,height=16cm]{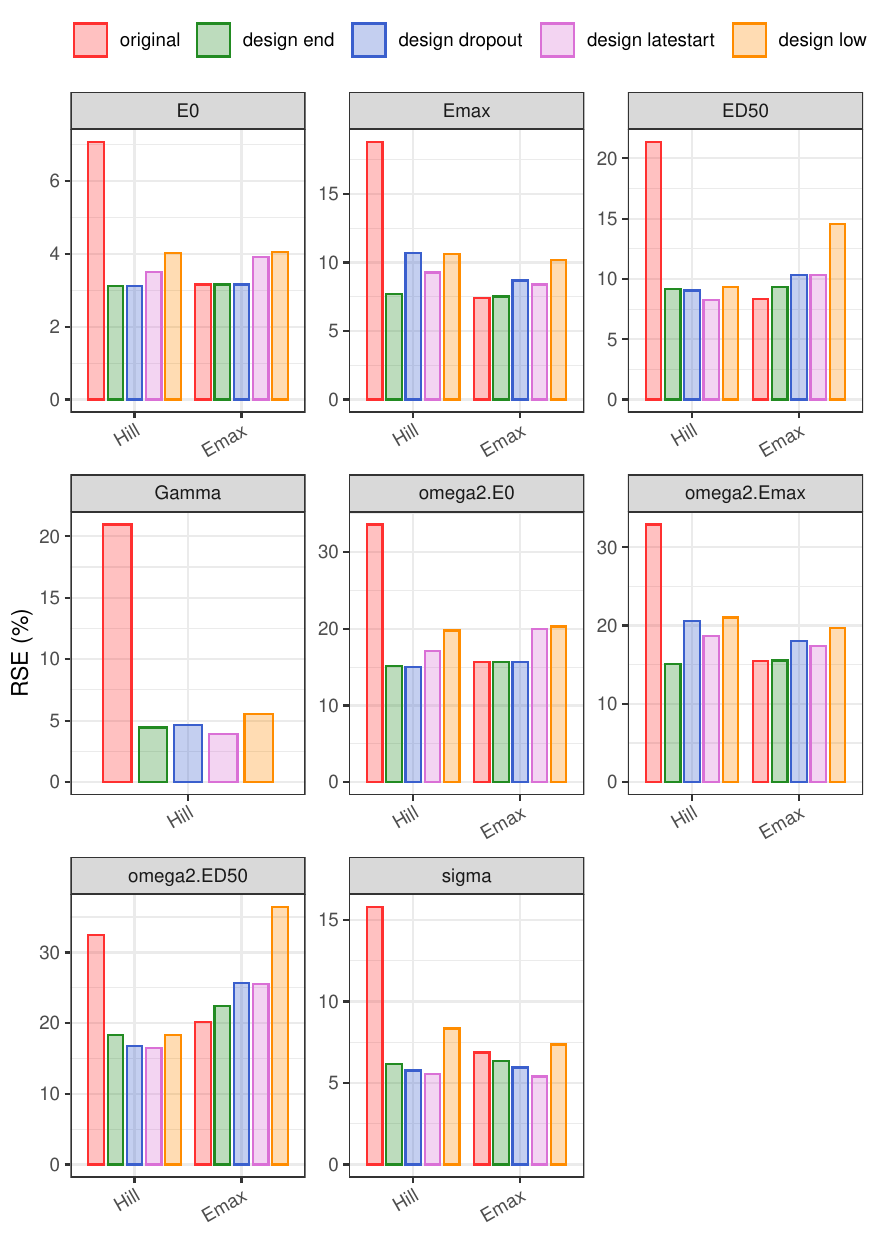}}
    \caption{\label{fig:unbalanced_RSE}Barplots of predicted RSE by PFIM 4.0, across the unbalanced designs compared to the original rich designs for each parameter}
\end{figure}

In Figure \ref{fig:unbalanced_RSE} we present the expected RSE \% calculated by the R package PFIM 4.0 for each parameter. The results for each unbalanced design is presented in a different colour and compared to the ones from the original rich designs (with red colour).

\clearpage
\newpage
\subsection{Coverage rates} \label{App_unbalanced_CR}

\begin{figure}[!ht]
    \centering
    \makebox{\includegraphics[scale=0.8]{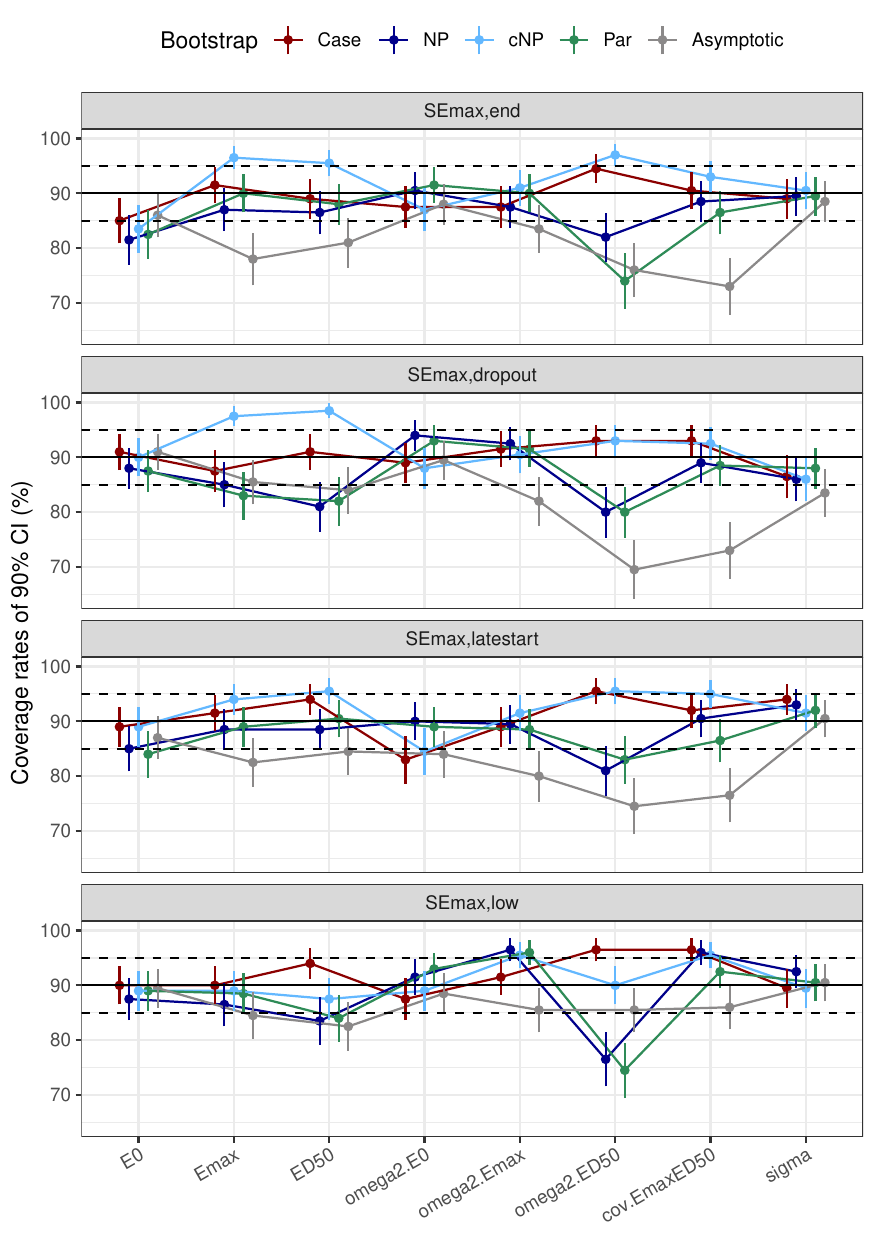}}
    \caption{\label{fig:CR90_unbalanced_Emax}Coverage rates of 90\% CI and errorbars of their MC uncertainty for the unbalanced designs for the $S_{{\rm E}_{max}}$. Dashed lines indicate a coverage of 85 and 95\%. The same scale is used on the Y-axis for the 4 graphs. The X-axis is jittered to avoid superimposing the different approaches.}
\par\kern -0.5cm
\end{figure}

\begin{figure}[!h]
    \centering
    \makebox{\includegraphics[scale=0.9]{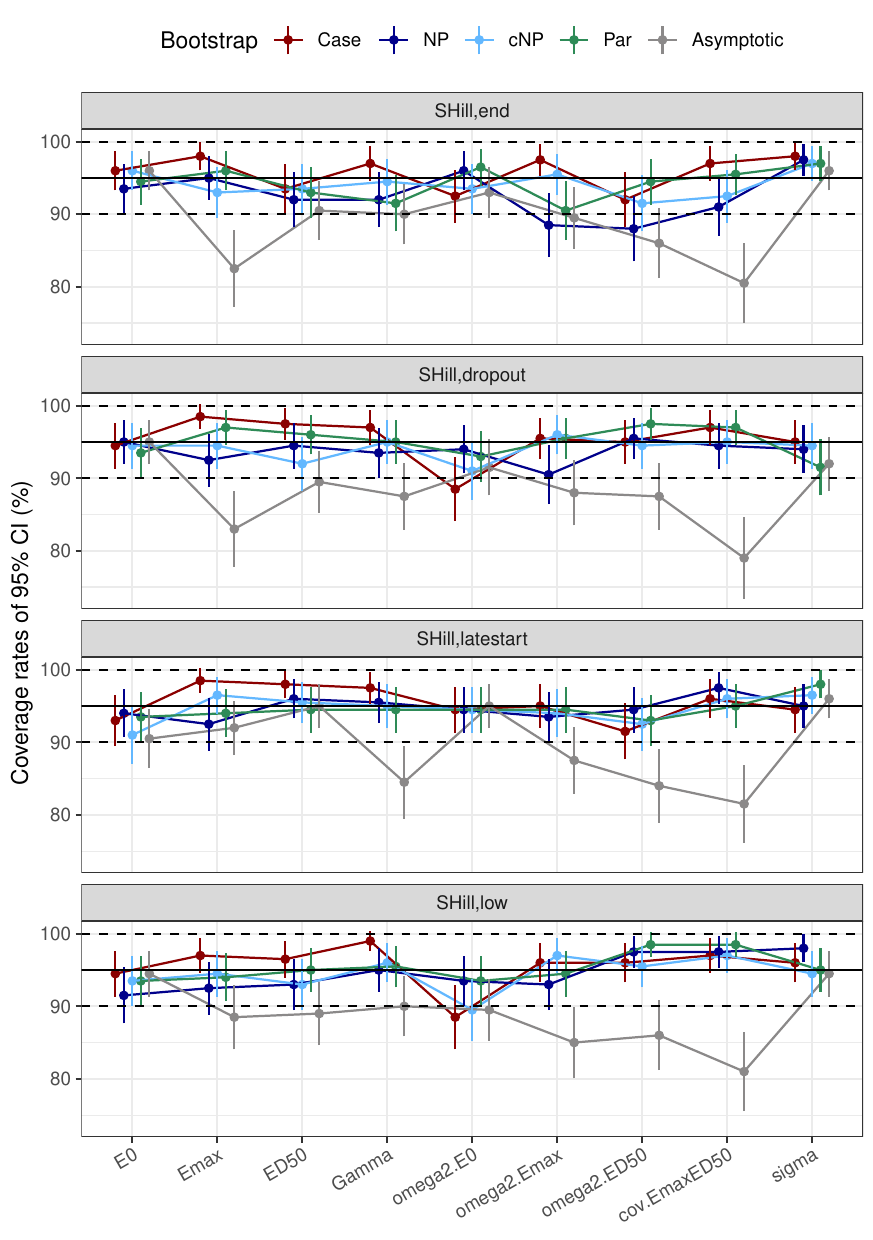}}
    \caption{\label{fig:CR95_unbalanced_Hill}Coverage rates of 95\% CI and errorbars of their MC uncertainty for the unbalanced designs for the $S_{Hill}$. Dashed lines indicate a coverage of 90 and 100\%. The same scale is used on the Y-axis for the 4 graphs. The X-axis is jittered to avoid superimposing the different approaches.}
\end{figure}

\begin{figure}[!h]
    \centering
    \makebox{\includegraphics[scale=0.9]{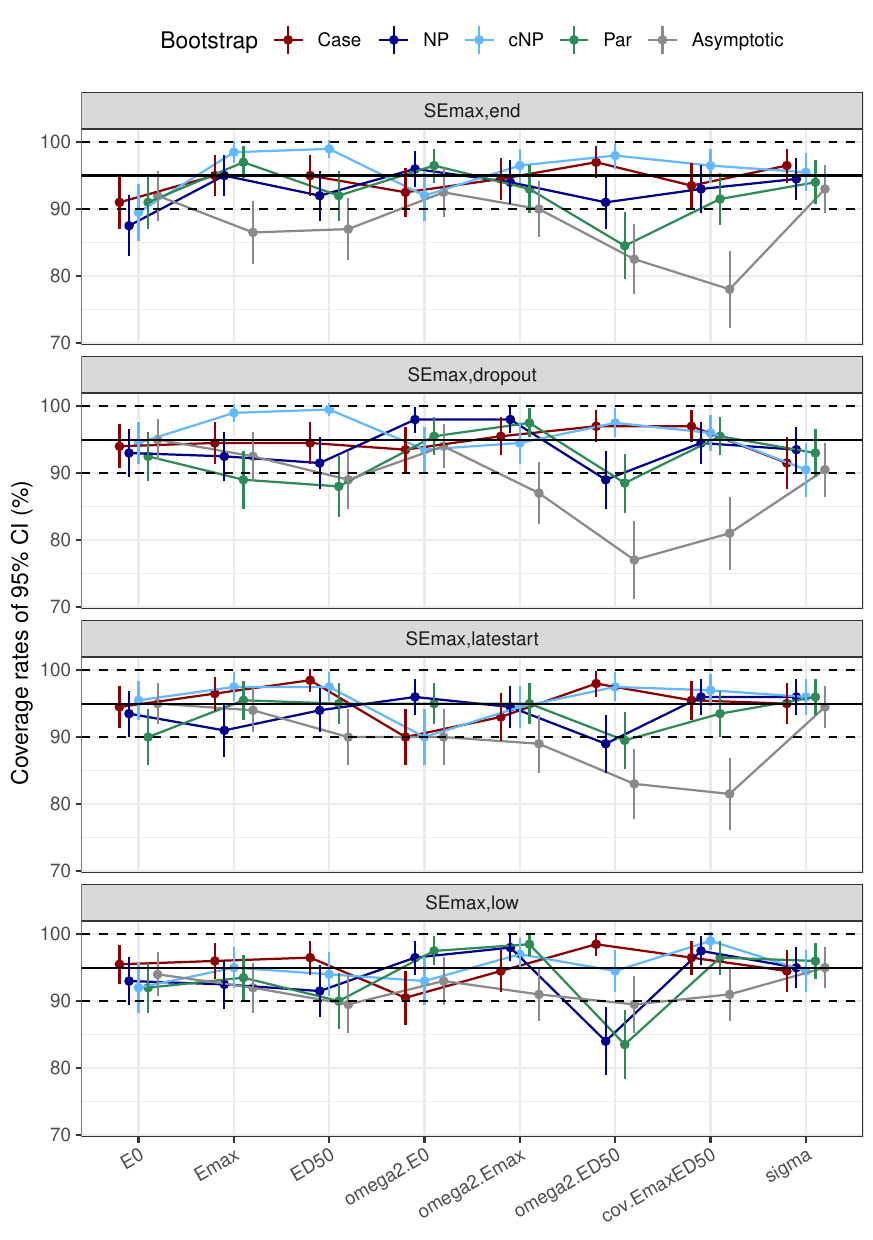}}
    \caption{\label{fig:CR95_unbalanced_Emax}Coverage rates of 95\% CI and errorbars of their MC uncertainty for the unbalanced designs for the $S_{{\rm E}_{max}}$. Dashed lines indicate a coverage of 90 and 100\%. The same scale is used on the Y-axis for the 4 graphs. The X-axis is jittered to avoid superimposing the different approaches.}
\end{figure}

Figure \ref{fig:CR90_unbalanced_Emax} shows the 90\% coverage rates obtained by all the bootstraps as well as the asymptotic method for the E$_{\rm max}$ scenarios in the four unbalanced designs, with similar findings as in Figure~\ref{fig:CR90_unbalanced_Hill}. In figures \ref{fig:CR95_unbalanced_Hill} and \ref{fig:CR95_unbalanced_Emax} we plot the coverage rates for the 95\% confidence intervals in the Hill and the E$_{\rm max}$ scenarios respectively. The results with the E$_{\rm max}$ were qualitatively similar to those obtained with a Hill model, except for the $S_{{\rm E}_{\rm max}, dropout}$ scenario where the cNP had wider CI than expected for E$_{\rm max}$ and ED$_{\rm 50}$ while NP and Par had narrower CI than expected for $\omega_{ED_{50}}$.

\par\kern -0.5cm

\clearpage
\newpage
\subsection{Bias} \label{App_unbalanced_Bias}

\begin{figure}[!h]
    \centering
    \makebox{\includegraphics[scale=0.8]{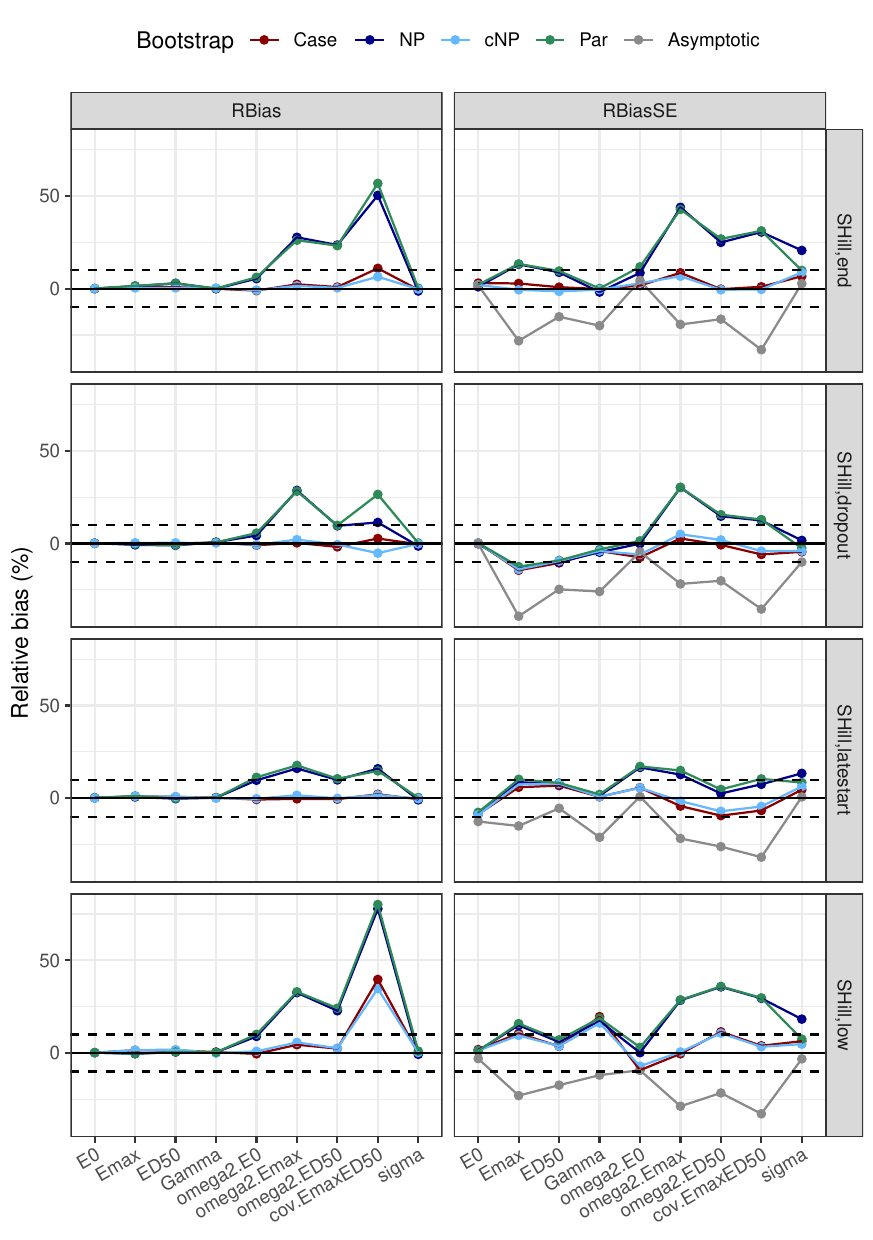}}
    \caption{\label{fig:Bias_unbalanced_Hill}Relative bias on parameters (left) and SE (right) for the original dosing scheme and the four unbalanced designs under the $S_{Hill}$. Dashed lines delineate absolute relative biases within 10\%. The same scale on the Y-axis is used}
\par \kern -0.5cm
\end{figure}

\begin{figure}[!ht]
    \centering
    \makebox{\includegraphics[scale=0.9]{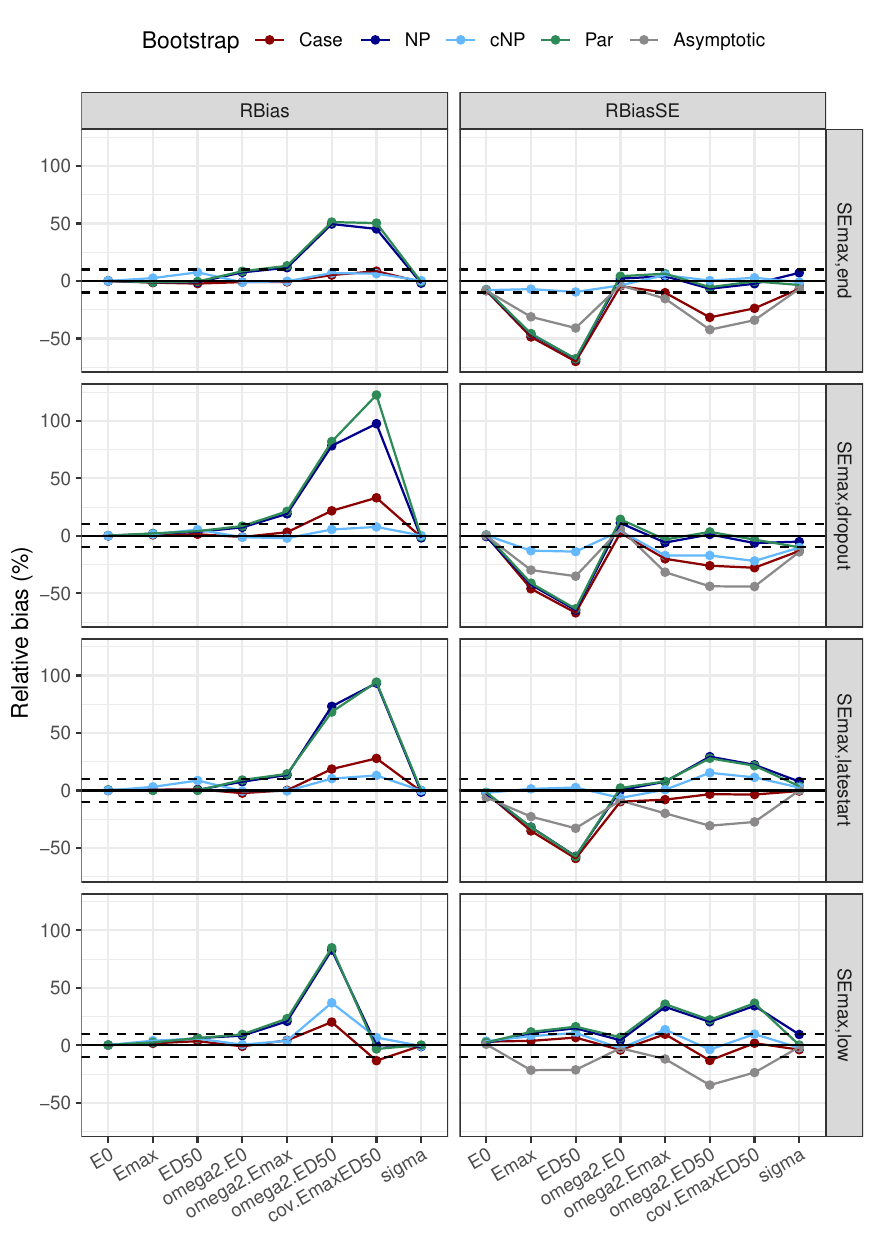}}
    \caption{\label{fig:Bias_unbalanced_Emax}Relative bias on parameters (left) and SE (right) for the original dosing scheme and the four unbalanced designs under the $S_{Emax}$. Dashed lines delineate absolute relative biases within 10\%. The same scale on the Y-axis is used}
\end{figure}

Figure \ref{fig:Bias_unbalanced_Hill} shows the bias on the parameters and their S.E. obtained by all the bootstraps as well as the asymptotic method for the Hill scenarios in the four unbalanced designs. In Figure \ref{fig:Bias_unbalanced_Emax} we plot the respective results for the E$_{\rm max}$ scenarios. 

\section{Appendix: Increased error} \label{App_error}

\subsection{Empirical SE} \label{App_error_SEemp}

\begin{figure}[!h]
    \centering
    \makebox{\includegraphics[width=13cm,height=16cm]{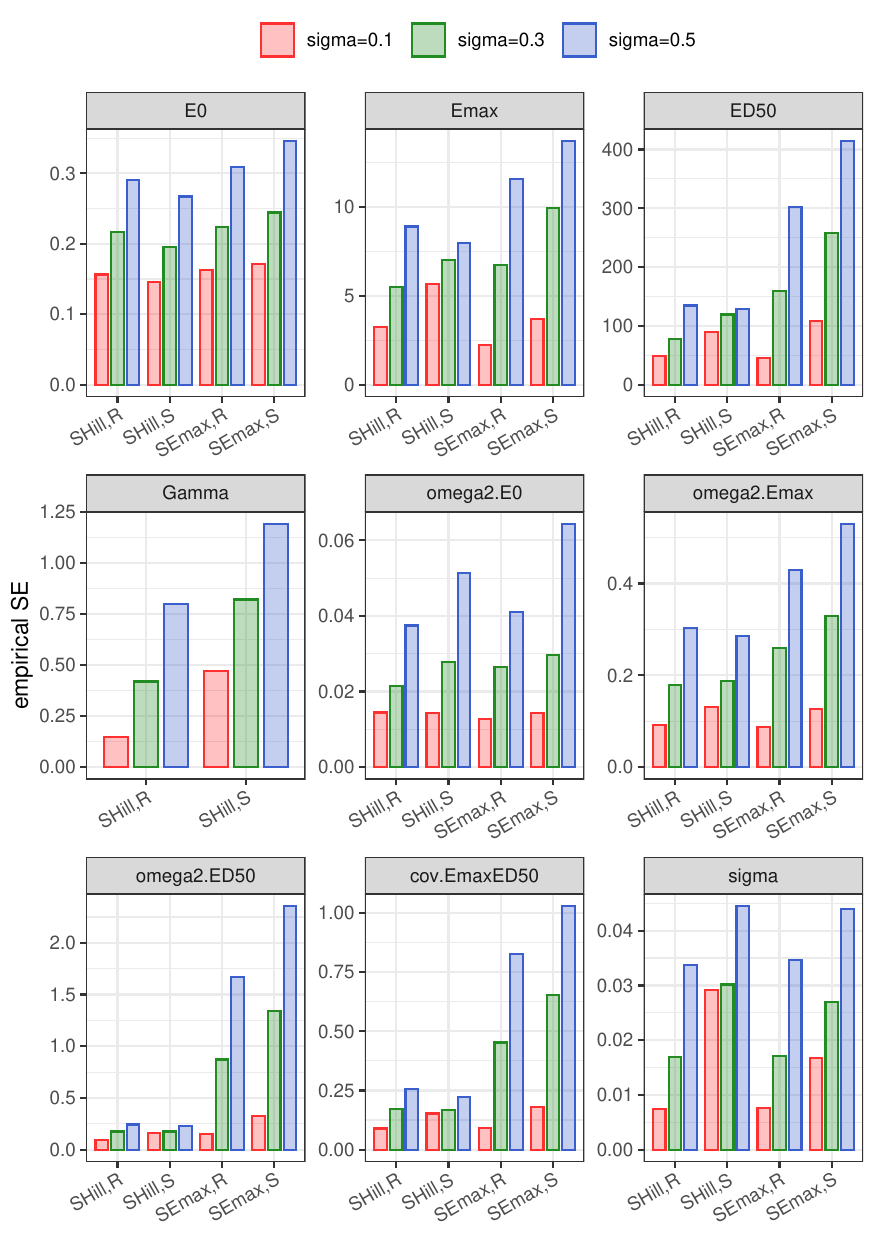}}
    \caption{\label{fig:error_empiricalSE} Barplots of emprical SE for the scenarios with increased coefficient compared across the four designs tested}
\end{figure}

In Figure \ref{fig:error_empiricalSE} we present the empirical SE for the scenarios with increased coefficient compared to the original designs with $\sigma=0.1$. Each graph window corresponds to a different parameter and the empirical SE for each scenario is presented with a bar of a different colour across the four designs tested.

\clearpage
\newpage

\subsection{Coverage rates} \label{App_error_CR}

\begin{figure}[!ht]
\par \kern -0.2cm
    \centering
    \makebox{\includegraphics[scale=0.75]{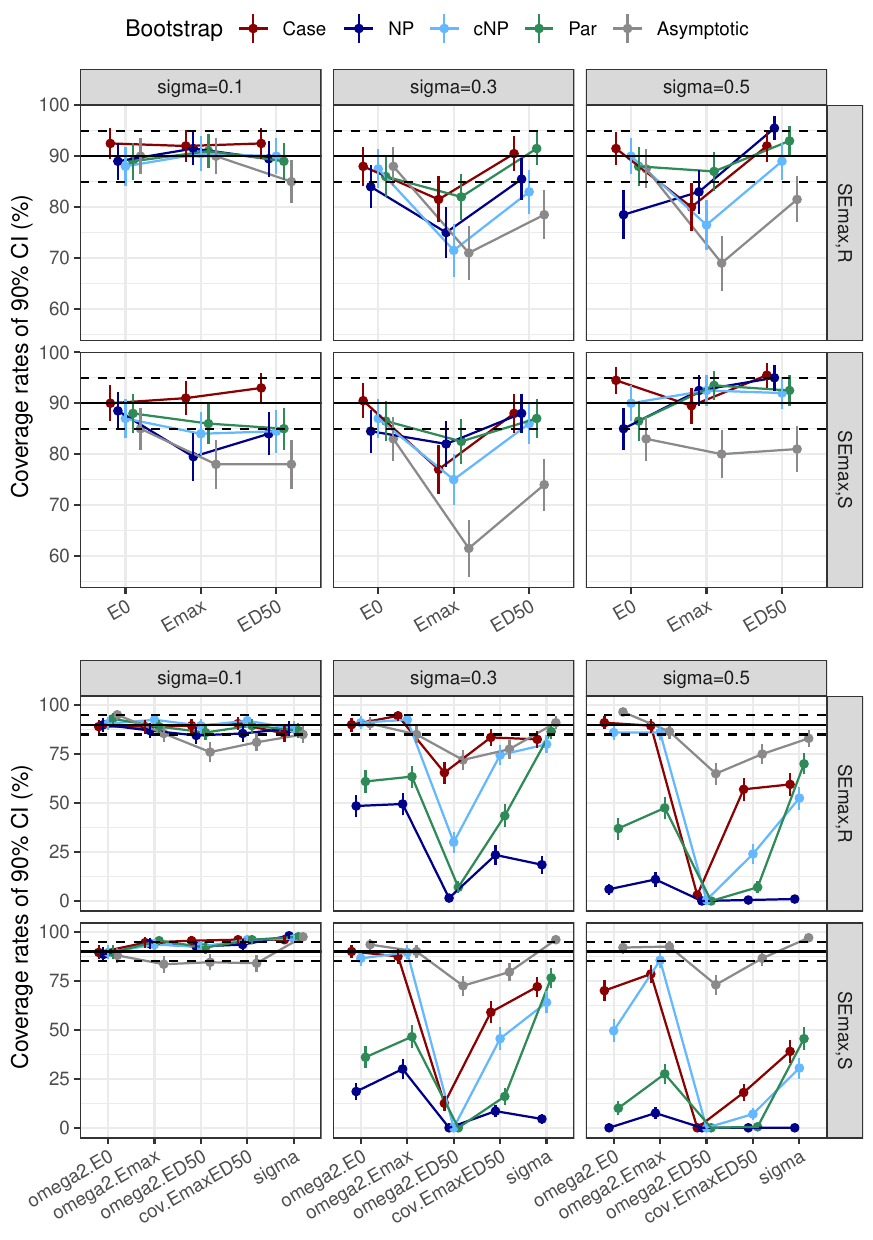}}
    \caption{\label{fig:CR90_error_Emax}Coverage rates of 90\% CI and errorbars of their MC uncertainty for the original design and the two designs with increased error coefficient $\sigma=0.3$ and $\sigma=0.5$ under the $S_{{\rm E}_{max}}$. Dashed lines indicate a coverage of 85 and 95\%. The same scale is used on the Y-axis for the 4 graphs. The X-axis is jittered to avoid superimposing the different approaches. At the first pair of lines we show the fixed effects and at the second the random effects. Each pair of lines has the same scale on the Y-axis}
\par \kern -0.5cm
\end{figure}

\begin{figure}[!ht]
    \centering
    \makebox{\includegraphics[scale=0.9]{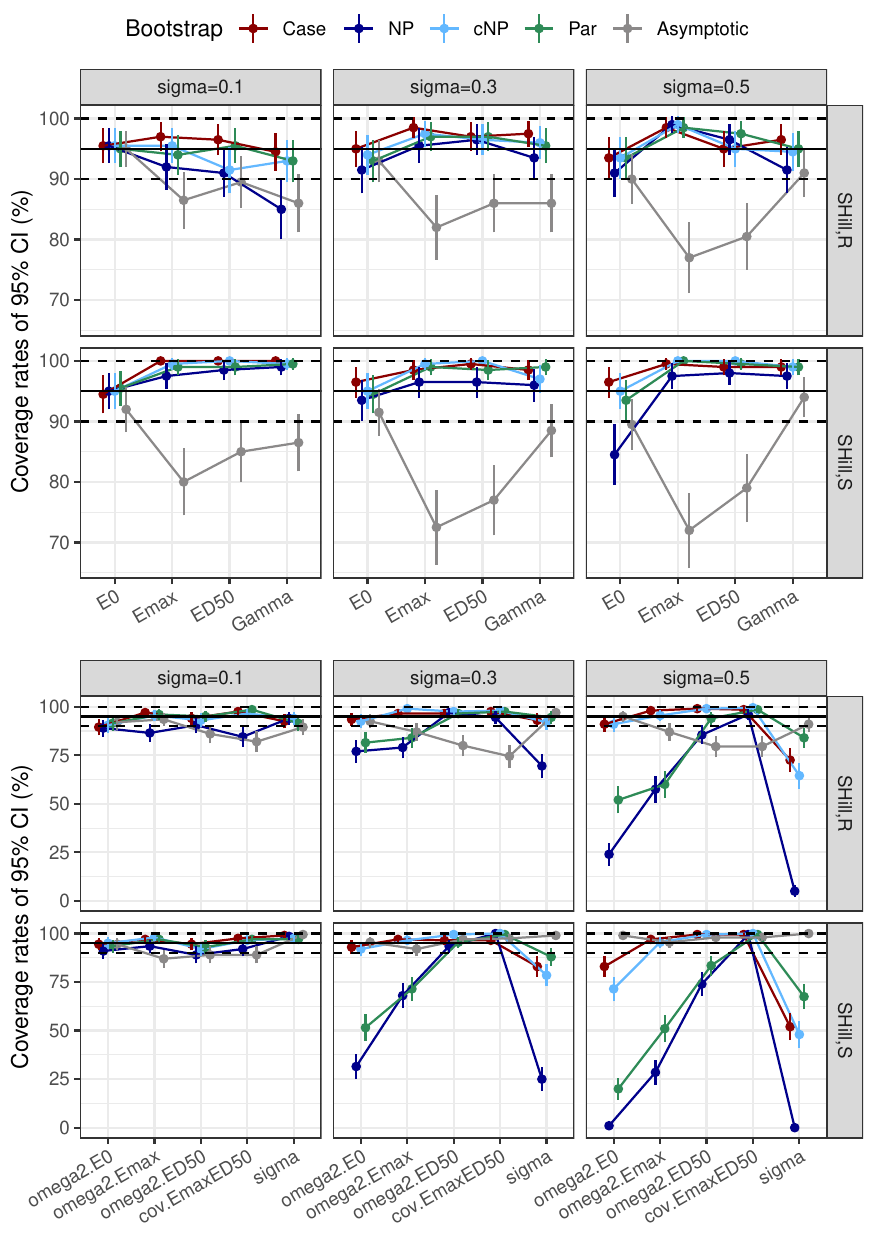}}
    \caption{\label{fig:CR95_error_Hill}Coverage rates of 95\% CI and errorbars of their MC uncertainty for the original design and the two designs with increased error coefficient $\sigma=0.3$ and $\sigma=0.5$ under the $S_{Hill}$. Dashed lines indicate a coverage of 90 and 100\%. The same scale is used on the Y-axis for the 4 graphs. The X-axis is jittered to avoid superimposing the different approaches. At the first pair of lines we show the fixed effects and at the second the random effects. Each pair of lines has the same scale on the Y-axis}
\end{figure}

\begin{figure}[!ht]
    \centering
    \makebox{\includegraphics[scale=0.9]{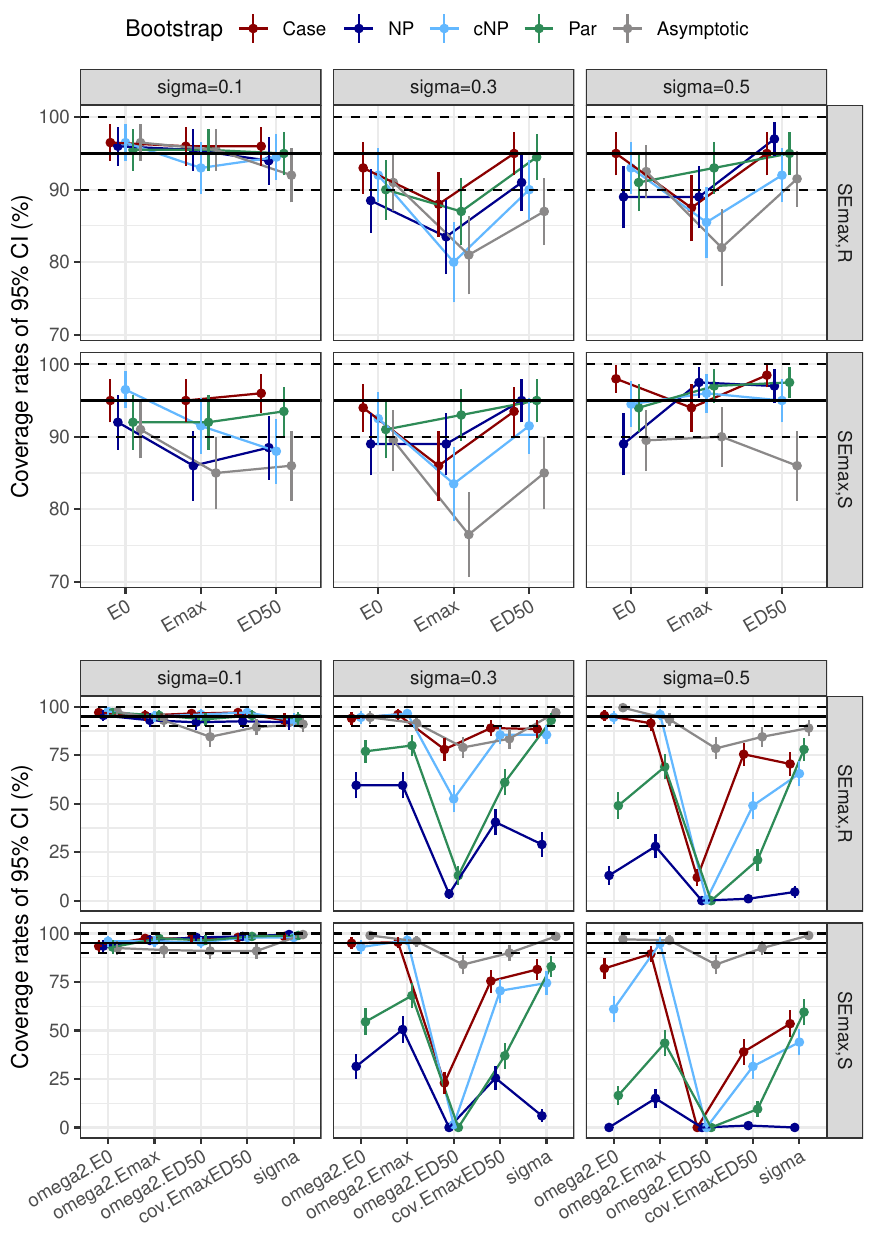}}
    \caption{\label{fig:CR95_error_Emax}Coverage rates of 95\% CI and errorbars of their MC uncertainty for the unbalanced designs. Dashed lines indicate a coverage of 90 and 100\%. The same scale is used on the Y-axis for the 4 graphs. The X-axis is jittered to avoid superimposing the different approaches.}
\end{figure}

Figure \ref{fig:CR90_error_Emax} shows the coverage rates for the 90\% CI obtained by all the bootstraps as well as the asymptotic method for the fixed effects (first pair of lines) and the random effects (second pair of lines) for the E$_{\rm max}$ scenarios in the original design and in the designs with increased error coefficient $\sigma$ at 0.3 and 0.5. In Figures \ref{fig:CR95_error_Hill} and \ref{fig:CR95_error_Emax} we present the 95\% coverage rates. All bootstraps show some bias as $\sigma$ increases, which could suggest that with high residual variability random and residual variability may become difficult to separate. However, there was no correlation between the estimates of variance terms and the estimates of $\sigma$ (data not shown) in our simulations to indicate this.

\par \kern -0.5cm

\clearpage

\clearpage
\newpage
\subsection{Bias} \label{App_error_Bias}

\begin{figure}[!ht]
    \centering
    \makebox{\includegraphics[scale=0.8]{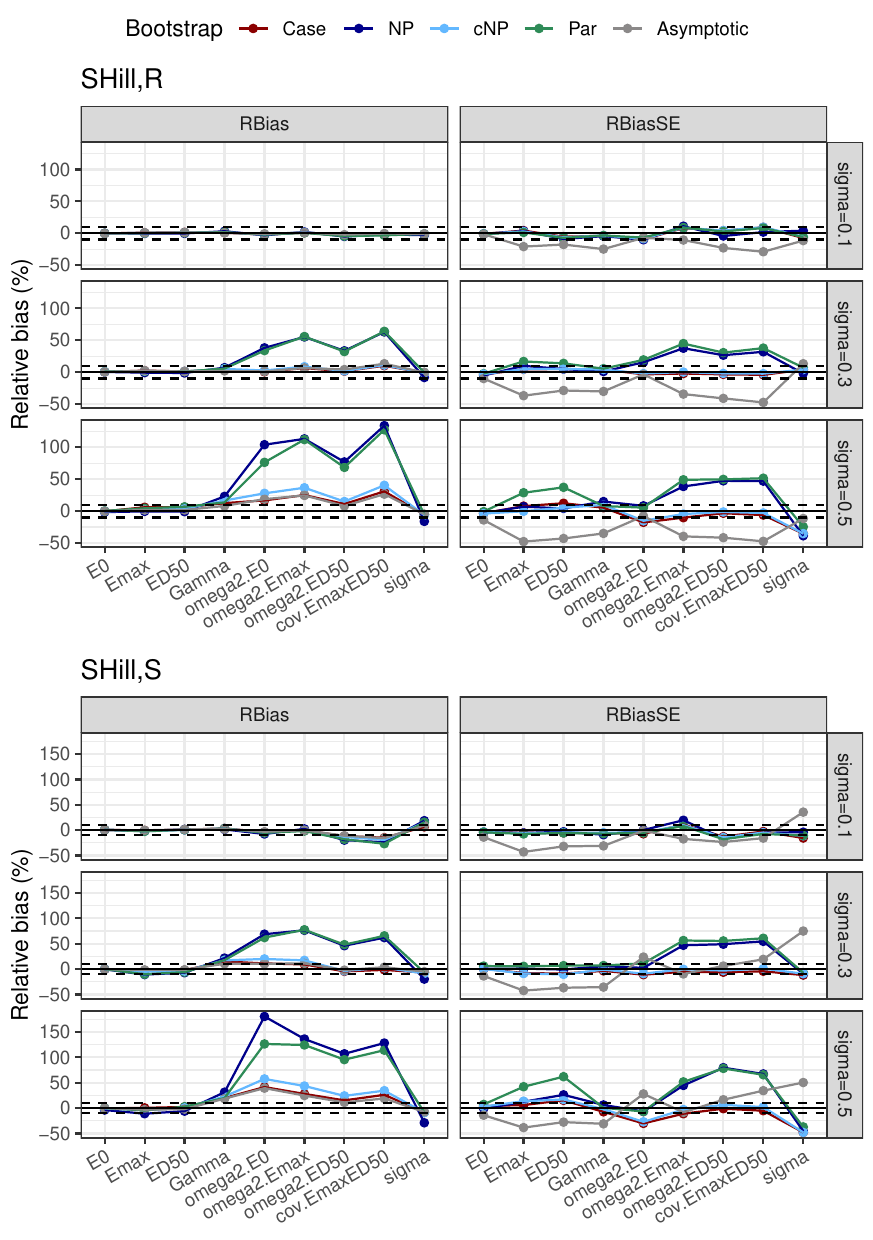}}
    \caption{\label{fig:Bias_error_Hill}Relative bias on parameters (up) and SE (down) for the original dosing scheme and the two scenarios with increased error coefficient $\sigma=0.3$ and $\sigma=0.5$ under the $S_{Hill}$. Dashed lines delineate absolute relative biases within 10\%. A fixed scale is used on the Y-axis}
\end{figure}

\begin{figure}[!ht]
    \centering
    \makebox{\includegraphics[scale=0.9]{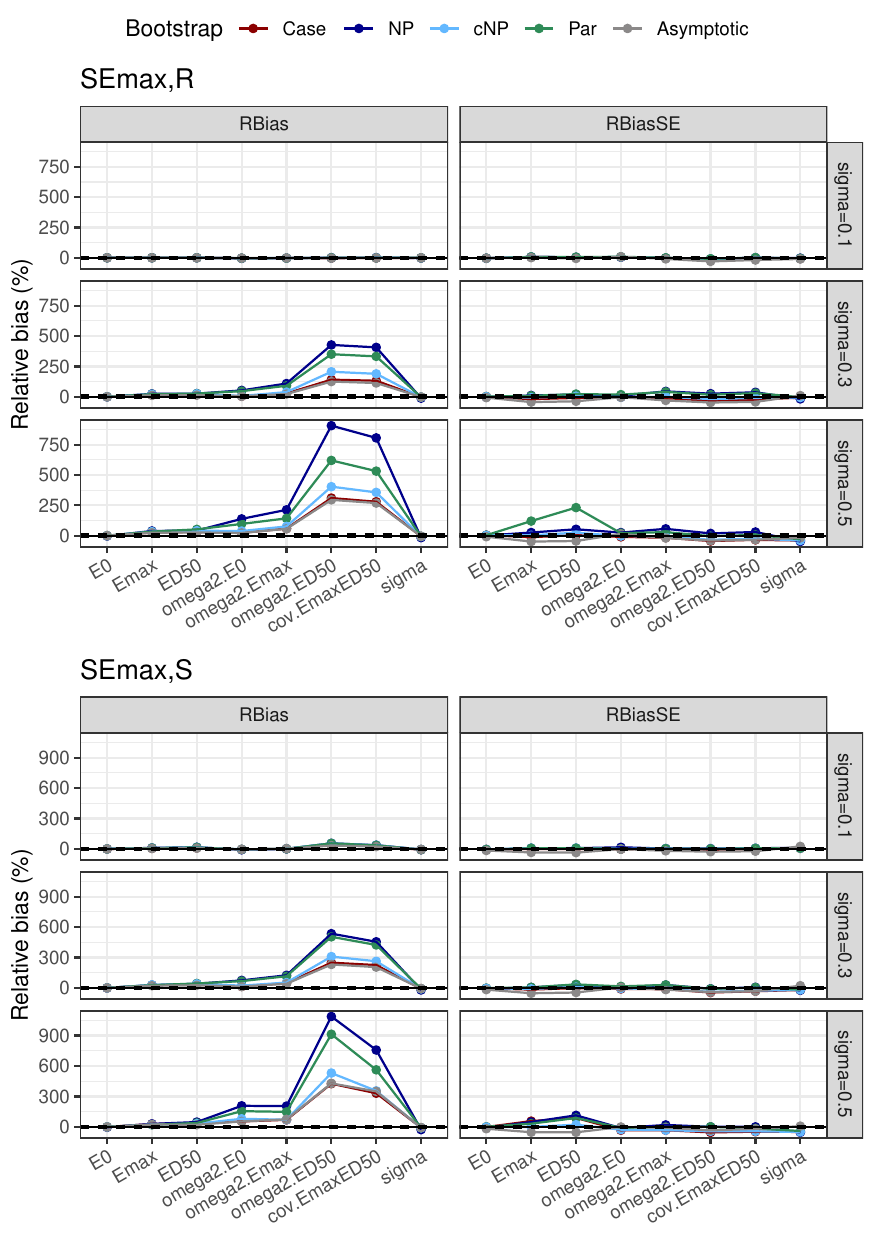}}
    \caption{\label{fig:Bias_error_Emax}Relative bias on parameters (up) and SE (down) for the original dosing scheme and the two scenarios with increased error coefficient $\sigma=0.3$ and $\sigma=0.5$ under the $S_{Emax}$. Dashed lines delineate absolute relative biases within 10\%. A fixed scale is used on the Y-axis}
\end{figure}

Figure \ref{fig:Bias_error_Hill} shows the bias on the parameters and their S.E. obtained by all the bootstraps as well as the asymptotic method for the Hill scenarios in the original design and in the designs with increased error coefficient $\sigma$ at 0.3 and 0.5. In Figure \ref{fig:Bias_error_Emax} we plot the respective results for the E$_{\rm max}$ scenarios.

\clearpage
\newpage

\end{document}